\documentclass[a4paper,oneside,10pt]{article}

\usepackage{epsfig,array,amsmath,amssymb,psfrag,graphicx,tensor,color}
\definecolor{c1}{RGB}{100, 0, 200}
\definecolor{blue1}{RGB}{0, 0, 230} 
\definecolor{red1}{RGB}{255,0,0}

\usepackage[colorlinks=true, linkcolor=blue1, citecolor=red, urlcolor=blue1]{hyperref}

\newcommand{\noi}{\noindent}
\newcommand{\beq}{\begin{equation}}
\newcommand{\eeq}{\end{equation}}
\newcommand{\bea}{\begin{eqnarray}}
\newcommand{\eea}{\end{eqnarray}}

\newcommand{\RR}{\mathbb{R}}

\newcommand{\mc}[1]{\mathcal{#1}}
\newcommand{\mb}[1]{\mathbf{#1}}

\newtheorem{thrm}{Theorem}
\newtheorem{prop}{Proposition}[section]

\makeatletter \@addtoreset{equation}{section} \makeatother 
\renewcommand{\theequation}{\arabic{section}.\arabic{equation}}

\begin{document}

\title{\bf On the relations between two construction methods for conserved currents of\\ differential equations}

\author{
G\'abor Zsolt T\'oth
\\[4mm] 
\small \textit{Wigner Research Centre for Physics,} \\
\small \textit{Konkoly-Thege Mikl\'os \'ut 29-33, 1121 Budapest, Hungary} \\
\small E-mail: \texttt{toth.gabor.zsolt@wigner.hu} \\
\date{}
}
\maketitle

\begin{abstract}
The relations between two construction methods (called multiplier and embedding methods)
for conserved currents of general systems of
ordinary or partial differential equations (DEs)
are investigated. Recent studies indicate that the multiplier method, which is a generalization of Noether's theorem, 
has significant advantages in comparison with the embedding method,
which uses adjoint-symmetry/symmetry pairs and is based on embedding 
the original system of DEs in a larger one that follows from a Lagrangian.
In particular, the multiplier method can generally give a wider range of conserved currents than the embedding method.
In this paper simple extended forms of general systems of DEs, obtained by treating 
parameters present in the equations or introduced into them as dependent variables, are studied.
A variant of a fundamental result on the connection between the embedding method and the action of symmetries on 
conserved currents that correspond to a multiplier is derived 
for the extended systems of DEs. 
Using this connection and by considering particular extensions that endow the extended DEs 
with scaling symmetry, it is shown that the embedding method 
becomes significantly stronger if it is also allowed to be applied to the extended forms of the 
original DEs.
It is also shown that up to equivalence the multipliers of an extended DE system contain
the parametric multipliers of the 
original system together with the derivatives of the corresponding conserved currents with respect to the parameters.
\end{abstract}  

\vspace{1cm}
\noi
 
\maketitle

\thispagestyle{empty}

\newpage



\section{Introduction}
\label{sec.intr}

For systems of (partial or ordinary) differential equations that have a Lagrangian formulation 
it is well-known that conserved currents can be obtained from variational symmetries
by Noether's theorem (see Theorem 5.58 and Definition 5.51 in \cite{Olver}). 
For non-variational differential equations (DEs) a generalization of Noether's theorem is the \emph{multiplier method},
in which the role of symmetries is taken over by conservation law multipliers \cite{Olver, BCA, lma, Vinogradov, AB1, AB2, Anco2}. 
Another method for finding conserved currents for a system of DEs that is not necessarily variational 
is to embed it in
a larger system of Euler--Lagrange equations in a certain way, and to apply Noether's theorem to the larger system 
\cite{Olver} (Exercise 5.37), \cite{Cav}-\cite{Ma}, \cite{Anco1}.
This method involves introducing additional dependent variables and
is not a generalization of Noether's theorem, as it does not reduce to Noether's theorem in its usual form
in the case of variational DEs. 
It also requires adjoint-symmetries in addition to symmetries,
if one aims to construct local conserved currents of the original DE system.
It is possible to formulate it without referring to Lagrangians, the Euler--Lagrange equations and Noether's theorem,
and in a large part of the literature, including a number of the references cited above, it is formulated in such a way.  
The connection between the Lagrangian and non-Lagrangian formulations 
was established and thoroughly discussed recently in \cite{Anco1}.
The main formula by which the conserved current is obtained in this method is called 
adjoint-symmetry/symmetry formula in \cite{Anco1}. 
For the Lagrangian approach the works \cite{Ibr1,Ibr5} 
should be noted as the starting point of intense further research.  

The multiplier method is well developed and is known to be suitable for finding, 
up to equivalence, the local conserved currents of 
a system of DEs if it satisfies certain mild regularity conditions \cite{BCA, Anco2}. 
Nevertheless, the second method, which will be called \emph{embedding method}, is also interesting, 
because it extends the Lagrangian formalism to arbitrary differential equations and 
has some other uses besides constructing local conserved currents (see the end of Section \ref{sec.emb}).
It also keeps the feature of Noether's theorem that it associates conserved currents with symmetries.
It is thus natural to ask what the relation between
the two methods is, and in particular whether the conserved currents given by the multiplier method can be reproduced by the embedding
method as well. These questions were investigated recently in \cite{Anco4, Anco1}.  

The embedding method
can be used to construct a local conserved current if an adjoint-symmetry and a symmetry of the system of  
DEs under consideration are given. Since a conservation law multiplier is also an adjoint-symmetry, 
the embedding method also gives a conserved current for a multiplier/symmetry pair. 
In \cite{Anco4, Anco1} it was shown that the latter current is equivalent with the current that is obtained by the action of the 
symmetry on the current that belongs to the multiplier, and on the basis of this result it was argued in \cite{Anco1} that generally
there can be conservation laws that the embedding method cannot reproduce, 
indicating that the embedding method is less powerful than the multiplier method.

The aim of the present paper is to continue the investigation of the two methods and the relations between them, 
and to show, in particular, that the embedding method becomes significantly stronger if certain extended forms
of the original system of differential equations are also taken into account.
These extended DE systems are obtained 
by promoting some constant parameters $g_l$ that are present in the original equations or 
are introduced into them to dependent variables and appending to the original equations the additional equations $\partial_\mu g_l=0$, 
which express the requirement that $g_l$ are constant.
The aim of extending a DE system in this way is to endow it with additional---primarily scaling---symmetries
that can be used in the embedding method.

The idea of extending DE systems in the way described above and thereby endowing them with scaling symmetry appeared 
recently in \cite{Anco2} (see Section 7 of \cite{Anco2}) as well,  
in relation to the problem of finding the conserved current that corresponds to a given multiplier.
The present paper is inspired to a large extent by these ideas and develops them further.

Regularity assumptions are often made on the DE systems in general investigations
of the various conserved current construction methods---see e.g.\ \cite{Anco2}. In this paper 
we assume only the differentiability of the DEs to sufficiently high orders and
we allow anticommuting (Grassmann algebra valued) dependent variables as well. 
This requires some changes in a few arguments in comparison with the literature.

The paper is organized as follows: In Section \ref{sec.prel} various notations and conventions are fixed.  
In Section \ref{sec.mthds} the multiplier method and the embedding method are recalled.  
Section \ref{sec.rel}, which is the central part of the paper, contains the discussion of the 
extension of the original system of DEs and the results that we obtained about applying 
the two conserved current construction methods to the extended DE systems. 
Section \ref{sec.concl} contains concluding remarks.
In Appendix \ref{sec.noether} Noether's theorem about the symmetries of Lagrangians and the corresponding conserved currents
is reviewed briefly for completeness.

\section{Preliminaries}
\label{sec.prel}

Instead of `dependent variables' we use the term \emph{fields} in the rest of the paper, with the applications of
continuous symmetries and conservation laws in field theory in mind.
Nevertheless, this is only a naming convention and does not imply any restriction on the types of the dependent and independent
variables and on the DEs that are considered.  

Indices will be used in two ways; often, in the same way as in the abstract index notation,
they will merely indicate that a quantity has multiple components 
and show what types of indices label the components,
but occasionally they will also label the particular components of multicomponent quantities.
The ranges of the indices will mostly be omitted, and the indices themselves will also be omitted in arguments of functions.
The Einstein summation convention will be understood to apply to 
pairs of identical indices consisting of an upper and a lower index.
Differentiation with respect to a variable that has indices yields a result
that has corresponding indices in the opposite (upper or lower) position.
For simplicity, the various indices will be assumed to take only finitely many values.

$u_i$ will denote a collection of fields or field components, assumed to be defined on a base manifold $M$.
$x^\mu$, $\mu=0,1,\dots, D$, where $D=\dim(M) -1$, will be coordinates covering some open domain $\Omega$ in $M$.
It is sufficient for the purposes of the paper to consider only a single coordinate patch,
and the behaviour of $u_i$ and other quantities under coordinate changes will not be important.
For simplicity, $u_i$ will be assumed to be real for all values of $i$, 
and complex fields are taken into account as two real fields.   
There would be no difficulty in allowing $u_i$ to be complex for some values of $i$, 
but it makes some of the formulas lengthier, since 
if $u_i$ is complex, then both $u_i$ and $u_i^*$ have to be included in them.
A field that has values in a manifold can be taken into account as an array of real field components, 
which are obtained by introducing coordinates in the target manifold. 
Sections of fibre bundles can also be taken into account as arrays of field components, which arise by choosing a trivialization
of the fibre bundle over $\Omega$ and introducing coordinates in the fiber
(it is assumed that $\Omega$ is chosen so that the fiber bundle can be trivialized over it).
A field configuration (or a configuration of $u_i$) will be understood to be the graph of $u_i$ over $\Omega$.

$u_i$ is allowed to be anticommuting (Grassmann algebra valued) for some values of $i$
in order to accommodate fermionic fields.
The following sign convention will be used for derivatives with respect to anticommuting variables: 
if $\theta$ is an anticommuting variable and $E$ is an expression of the form 
$X_1\theta X_2$, then $\frac{\partial E}{\partial \theta}= (-1)^n X_1 X_2$, where $n=0$ if $X_2$ is even and $n=1$ if 
$X_2$ is odd.

The square bracket notation $f[u]$, where $u_i$ are fields and the index $i$ is omitted,
will be used to indicate that $f$ is a \emph{local function} of $u_i$, 
which means that it is a function of $x^\mu$, $u_i(x)$ and finitely many derivatives of $u_i(x)$. 
Local functions are also called differential functions in the literature (see \cite{Olver}, for example).
Total derivatives with respect to $x^\mu$ will be denoted by $D_\mu$.
The collection of all $k$-th order derivatives of $u_i$, for example, with respect to $x^\mu$,
i.e.\ $\partial_{\mu_1 \mu_2\dots\mu_k}u_i$,
will be denoted by $\partial^k u$ when an index-free notation is needed. 

We shall consider general systems of differential equations
\beq
\label{eq.de}
F^a[u](x) \equiv F^a(x,u(x),\partial u(x),\partial^2 u(x),\dots,\partial^k u(x))=0\, .
\eeq
The index $a$ labeling the equations is generally not related to the index $i$ that labels the fields, and 
$F^a$ is assumed to have definite commutation properties, i.e.\ it is either even or odd, for each value of $a$. 
We also assume that $F^a$ is differentiable as many times as necessary
with respect to $x^\mu$, $u_i$, $\partial_\mu u_i$, $\partial_{\mu\nu} u_i$, \dots 
(regarded as independent variables), 
but further assumptions on $F^a$ are not made, unless explicitly stated. 
In particular, the regularity of (\ref{eq.de}) in the sense of \cite{Anco2} is not assumed.
In the rest of the paper, $F^a=0$ will be understood to mean that $u_i$ satisfies (\ref{eq.de}).

Let $L[u]$ be a local function of $u_i$. The \emph{Euler--Lagrange derivative} of $L$ with respect to $u_i$ is
\beq
\label{eq.el1}
\mb{E}^i(L)[u]=
\frac{\delta L}{\delta u_i} =
\frac{\partial L}{\partial u_i}-D_\mu\frac{\partial L}{\partial (\partial_\mu u_i)}
+D_{\mu\nu}\frac{\partial L}{\partial (\partial_{\mu\nu} u_i)}-
D_{\mu\nu\lambda}\frac{\partial L}{\partial (\partial_{\mu\nu\lambda} u_i)}+
\dots\, .
\eeq
The DEs $\mb{E}^i(L)[u]=0$
for $u_i$ are called the \emph{Euler--Lagrange equations} corresponding to $L$, 
which is called the \emph{Lagrangian density function} for (\ref{eq.el1}). 
Lagrangians will be assumed to be even, because the Lagrangians occurring in physics are usually even,
and because the auxiliary Lagrangian (to be introduced in Section \ref{sec.emb}), 
which is the most important type of Lagrangian in this paper, is also even.

A \emph{conserved current} for the system (\ref{eq.de}) is a current $J^\mu$ for which the \emph{conservation law} 
$D_\mu J^\mu=0$ holds if $F^a=0$.
A \emph{local current} is a current that is a local function of $u_i$. 
In this paper we focus mainly on \emph{local conserved currents} that are even.
The transformation properties of the currents under general coordinate changes will not be relevant. 
A current is \emph{identically conserved} if it is conserved for arbitrary $u_i$.
Two local conserved currents $J^\mu_1$ and $J^\mu_2$ are \emph{equivalent} (\cite{Olver}, Section 4.3) if   
$J^\mu_1-J^\mu_2=\bar{J}^\mu + \hat{J}^\mu$, 
where $\bar{J}^\mu$ and $\hat{J}^\mu$ are local currents, $\bar{J}^\mu$ is identically conserved
and $\hat{J}^\mu=0$ if $F^a=0$.

Although the definition above for the equivalence of currents is reasonable,
it should be noted that even if two currents are equivalent according to this definition, 
they may be physically inequivalent if $\bar{J}^\mu\ne 0$.
For example, the electric current is physically nontrivial in spite of being equivalent
with the zero current in virtue of Maxwell's equations.

A \emph{one-parameter family of transformations} of the fields $u_i$ 
is given by a mapping $T_i\{\varsigma,u\}$, where the independent variables of $T_i$ are the
parameter $\varsigma$ ($\varsigma \in \RR$) and the field configuration $u_i$. 
For given $\varsigma$ and $u_i$,
$T_i\{\varsigma,u\}$ are the transformed fields.
The curly brackets $\{\, \}$ are used here to indicate that 
$T_i\{\varsigma,u\}$ can be a general function of the field configuration, i.e.\
it is not necessarily a local function of $u_i$.
$T_i\{\varsigma,u\}(x)$, $x^\mu\in \Omega$, is assumed to be defined for any $\varsigma$ 
in some open interval, which contains $0$ and
may depend on $x^\mu$ and on the configuration of $u_i$. 
It is also assumed that $T_i\{0,u\}=u_i$.
$T_i\{\varsigma,u\}$ is not required to have a group property. 

After linearization in the parameter, $u_i\to T_i\{\varsigma,u\}$ becomes
\beq
\label{eq.tr1}
u_i\to u_i+\varsigma\,\delta u_i \{ u \}\, ,
\eeq
where $\delta u_i\{u\} = \frac{d  T_i\{\varsigma,u\}}{d\varsigma}|_{\varsigma=0}$. 
$\delta u_i$ has the same commutation character 
as $u_i$ for all values of $i$.
In the present paper only the linearized transformations (\ref{eq.tr1}) are needed, 
and it would be sufficient to consider only such $T_i\{\varsigma,u\}$ 
that are exactly linear in $\varsigma$.
$\delta u_i$ will also be assumed to be a local function of $u_i$.
Comparing with the definitions of \cite{Olver} and \cite{Anco2}, it can be said that we consider infinitesimal transformations 
in evolutionary (or characteristic) form (see Section 5.1 of \cite{Olver} or Section 5 of \cite{Anco2}),
and $\delta u_i$ is the \emph{characteristic} of the infinitesimal transformation (\ref{eq.tr1}).
For brevity, $\delta$ with a subscript will sometimes be used to refer to $u_i\to u_i+\varsigma\,\delta u_i \{ u \}$.
The \emph{first order variation}, under the action of $u_i\to T_i\{\varsigma,u\}$,
of a quantity $f\{u\}$ that depends on the field configuration is  
$\delta f \{ u \} = \frac{df\{u+\varsigma\delta u \{ u\} \} }{d\varsigma}|_{\varsigma=0}$.
If $f$ is a local function, then 
$\delta f = \frac{\partial f}{\partial u_i}\delta u_i 
+ \frac{\partial f}{\partial (\partial_\mu u_i)}D_\mu\delta u_i 
+ \frac{\partial f}{\partial (\partial_{\mu\nu} u_i)}D_{\mu\nu}\delta u_i +\dots $.
$f\{u\}$ is called \emph{homogeneous of weight (or order) $s$} 
under the action of (\ref{eq.tr1}) if $\delta f = s f$ with some constant $s$.

\section{The two conserved current construction methods}
\label{sec.mthds}

In this section we recall briefly the two methods mentioned in the Introduction for constructing conserved currents 
for systems of differential equations. Useful references about these methods are \cite{Olver, BCA, Anco2, Ibr4, SF, Anco1}. 
An extension of the embedding method and of Noether's theorem to boundary conditions can be found in \cite{t}.

\subsection{The conservation law multiplier method}
\label{sec.clm}

Let $q_a[u]$ be a local function for which 
\beq
\label{eq.mult}
F^a q_a = D_\mu J^\mu
\eeq
holds with some local current $J^\mu[u]$ for any configuration of $u_i$. 
For any value of $a$, $q_a$ is assumed to have the same commutation properties
as $F^a$.
Such a $q_a$ is called a \emph{conservation law multiplier}
for $F^a$ and $J^\mu$, and it is obvious from (\ref{eq.mult}) that $J^\mu$ is conserved for all solutions of (\ref{eq.de}).

In virtue of (\ref{eq.mult}), $q_a[u]$ is a conservation law multiplier if and only if 
the equation $\frac{\delta (F^a q_a)}{\delta u_i}=0$, called \emph{multiplier determining equation},
holds for arbitrary configurations of $u_i$.
This equation can be used to find multipliers for $F^a$, and then for any multiplier
it is a further problem to find the $J^\mu[u]$ that satisfies (\ref{eq.mult}).
Various methods for solving the latter problem are known \cite{Anco2}. 
From (\ref{eq.mult}) it is clear that a conserved current corresponding to a multiplier is determined up to adding an
identically conserved current.
If $F^a$ satisfies certain regularity conditions, then all local conserved currents, 
up to equivalence, correspond to a multiplier \cite{Anco2}.

As (\ref{eq.n1}) shows, any characteristic $\delta u_i$ of a symmetry of a Lagrangian density function 
is a conservation law multiplier for the Euler--Lagrange equations and vice versa, 
and the Noether current, multiplied by $-1$, is the current corresponding to $\delta u_i$ as multiplier,
therefore the multiplier method is a generalization of Noether's theorem to arbitrary DEs.
On the other hand, the characteristic of an on-shell symmetry of a Lagrangian is not necessarily a multiplier, 
thus the multiplier method is not a generalization of the on-shell symmetry version of Noether's theorem
(which is introduced in Appendix \ref{sec.oss}).

\subsection{Embedding of differential equations in a system of Euler--Lagrange equations}
\label{sec.emb}

The DE system (\ref{eq.de}) can be embedded in a system of Euler--Lagrange equations by adding to the set of fields   
a set of \emph{auxiliary fields} $\rho_a$, which have the same commutation properties as $F^a$, 
and taking the \emph{auxiliary Lagrangian density function} (also called formal Lagrangian)
\beq
\label{eq.l}
\hat{L}[u,\rho]=F^a[u]\rho_a\, .
\eeq
The Euler--Lagrange equations following from (\ref{eq.l}) for $\rho_a$ are just (\ref{eq.de}), and the Euler--Lagrange equations for $u_i$, 
\beq
\mb{E}^i(\hat{L})[u,\rho] =
\frac{\delta \hat{L}}{\delta u_i}
=\frac{\partial (F^a\rho_a)}{\partial u_i}-D_\mu\frac{\partial (F^a\rho_a)}{\partial (\partial_\mu u_i)}
+D_{\mu\nu}\frac{\partial (F^a\rho_a)}{\partial (\partial_{\mu\nu}u_i)}-
D_{\mu\nu\lambda}\frac{\partial (F^a\rho_a)}{\partial (\partial_{\mu\nu\lambda}u_i)}+
\dots = 0\, ,
\label{eq.elaux}
\eeq
constitute a further
set of equations, which are homogeneous linear in $\rho_a$.
As equations for $\rho_a$, (\ref{eq.elaux}) are also known as the
adjoint linearization of (\ref{eq.de}).
The complete set of Euler--Lagrange equations is satisfied if $u_i$ satisfy (\ref{eq.de}) and $\rho_a=0$, therefore
the Lagrangian system defined by (\ref{eq.l}) indeed properly contains (\ref{eq.de}).
If (\ref{eq.de}) are linear equations, 
then (\ref{eq.elaux}) are also linear and do not contain $u_i$ and their derivatives.

After embedding (\ref{eq.de}) in the Lagrangian system specified by (\ref{eq.l}), one can try to find symmetries of $\hat{L}$, 
and then the associated conserved currents can be found using the Noether construction described in Appendix \ref{sec.noether}.
In particular, if (\ref{eq.de}) has a symmetry, then $\hat{L}$ also has a corresponding on-shell symmetry, as described below.

A one-parameter transformation $u_i\to T_i\{\varsigma,u\}$ is a called a \emph{symmetry of the DE system (\ref{eq.de})}, if 
\beq
\label{eq.sc}
\delta F^a[u] = \frac{dF^a[u+\varsigma\,\delta u[u]]}{d\varsigma}|_{\varsigma=0}=0
\eeq
holds for any solution of (\ref{eq.de}). 
This symmetry condition is the infinitesimal form of the requirement that a symmetry is a transformation that maps a solution of (\ref{eq.de}) into another solution. 
Although it would be more appropriate to use the term `infinitesimal symmetry' instead of `symmetry' in this definition,
we consider only infinitesimal symmetries in this paper, therefore the word `infinitesimal' is omitted.
Regarding $\delta u_i$ as independent fields,
$\frac{dF^a[u+\varsigma\delta u]}{d\varsigma}|_{\varsigma=0}$
is just the \emph{linearization of $F^a$ around $u_i$}, 
thus if $u_i\to T_i\{\varsigma,u\}$ is a symmetry, then $\delta u_i[u]$ 
is a solution of the linearization of (\ref{eq.de}) around $u_i$ if $u_i$ is a solution of (\ref{eq.de}). 

If $u_i\to u_i+\varsigma\,\delta u_i$ is a symmetry of (\ref{eq.de}), then
$\delta \hat{L}=F^a\delta \rho_a+\delta F^a \rho_a$ is clearly zero if $F^a=0$, for any choice of $\delta\rho_a$.
This means that $u_i\to u_i+\varsigma\,\delta u_i$, $\rho_a\to\rho_a+\varsigma\,\delta\rho_a$
is an on-shell symmetry
(see Appendix \ref{sec.oss})
of $\hat{L}$ with $K^\mu=0$ and with arbitrary $\delta\rho_a$.
Since $K^\mu=0$, the associated Noether current is $j^\mu$ (see (\ref{eq.elc}) for the definition of $j^\mu$). Explicitly,
\bea
j^\mu & = &  \frac{\partial (F^a\rho_a)}{\partial (\partial_\mu u_i)}\delta u_i+
\biggl(\frac{\partial (F^a\rho_a)}{\partial (\partial_{\mu\nu} u_i)}D_\nu \delta u_i
-D_\nu\frac{\partial (F^a\rho_a)}{\partial (\partial_{\mu\nu}u_i)} \delta u_i\biggr)\nonumber\\
&& + \biggl(\frac{\partial (F^a\rho_a)}{\partial (\partial_{\mu\nu\lambda}u_i)}D_{\nu\lambda}\delta u_i
-D_\nu\frac{\partial (F^a\rho_a)}{\partial (D_{\mu\nu\lambda}u_i)}D_\lambda\delta u_i
+D_{\nu\lambda}\frac{\partial (F^a\rho_a)}{\partial (\partial_{\mu\nu\lambda}u_i)}\delta u_i\biggr)+\dots\, .
\label{eq.elc2}
\eea
$j^\mu$ is conserved if $u_i$ satisfies (\ref{eq.de}) and $\rho_a$ satisfies the auxiliary equations (\ref{eq.elaux}). 
Since $\hat{L}$ does not depend on the derivatives of $\rho_a$, 
$j^\mu$ does not depend on the choice of $\delta\rho_a$.  
$j^\mu$ is homogeneous linear in $\rho_a$, 
therefore it is necessary to find nonzero solutions of (\ref{eq.elaux}) for $\rho_a$ in order
to obtain nonzero $j^\mu$. 

$j^\mu$ becomes a local conserved current of (\ref{eq.de}) if 
$\rho_a$ is a local function of $u_i$ and satisfies (\ref{eq.elaux}) whenever $F^a=0$. 
A $\rho_a[u]$ with these properties is called an \emph{adjoint-symmetry}. 
Here it should be noted, to avoid confusion,
that the differentiations with respect to $\partial_{\mu\nu\dots}u_i$
in (\ref{eq.elaux}) and in (\ref{eq.elc2}) do not apply to $\rho_a[u]$. 
Nevertheless, it is clear from (\ref{eq.elc2}) that
if one applies the differentiations with respect to $\partial_{\mu\nu\dots}u_i$ in (\ref{eq.elc2}) to $\rho_a[u]$ as well, 
then the resulting current differs only by a current that is zero if $F^a=0$. Similarly, (\ref{eq.elaux}) would change 
only by a quantity that is zero when $F^a=0$, therefore the adjoint-symmetry property of $\rho_a[u]$ would not 
be affected. 
The strictly self-adjoint, quasi self-adjoint, weak self-adjoint and 
nonlinear self-adjoint differential equations \cite{Ibr1}-\cite{Gand1}, \cite{SF} are special types of differential equations admitting an adjoint-symmetry.

The above construction of conserved currents can be summarized as follows:
{\linespread{1.2}
\begin{thrm}
\label{th.emb1}
{\rm (i)} If $\delta u_i[u]$ is the characteristic of a symmetry of 
(\ref{eq.de}) and $\rho_a[u]$ is an adjoint-symmetry of (\ref{eq.de}), then $j^\mu$ is a local conserved current of (\ref{eq.de}).
{\rm (ii)} If $\delta u_i[u]$ is the characteristic of a symmetry of 
(\ref{eq.de}) and $\delta\rho_a[u,\rho]$ is arbitrary, then 
$\{ \delta u_i, \delta\rho_a \}$ is the characteristic of an on-shell symmetry of $\hat{L}[u,\rho]$ with $K^\mu=0$, 
and the corresponding Noether current of the Euler--Lagrange equations (\ref{eq.de}), (\ref{eq.elaux})
is $j^\mu$, which does not depend on $\delta\rho_a[u,\rho]$.
\end{thrm} }
Here and in the subsequent parts of the paper the brackets $\{ \}$ are used for collecting pieces of quantities that have multiple components,
i.e., for joining arrays.
Note that in the first part of the theorem $j^\mu$ is a local function of $u_i$, whereas in the second part 
it is a local function of $\{ u_i, \rho_a \}$, and the first part follows from the second part.
The two parts of Theorem \ref{th.emb1} can thus be regarded as two closely related constructions.

In the preceding arguments it was shown that a symmetry of (\ref{eq.de})
implies an on-shell symmetry of $\hat{L}$, 
and this was sufficient to find the Noether current $j^\mu$, without further assumptions on $F^a$. 
Since $j^\mu$ was obtained by the on-shell symmetry version of Noether's theorem, 
a multiplier to which $j^\mu$ corresponds was not found. 
In this respect our presentation differs from the literature, where 
relatively mild regularity assumptions on $F^a$ are made, under which it is possible
to specify $\delta\rho_a$ and $K^\mu$ so that the on-shell symmetry becomes a complete symmetry of $\hat{L}$
and $K^\mu=0$ continues to hold on the solutions of (\ref{eq.de})
(see Section 2.2 of \cite{Anco1} and \cite{Ibr1}). 
Nevertheless, the latter result will not be needed in the present paper. 

There is another construction, which is closely related to the one described in Theorem \ref{th.emb1}
and is worth mentioning:
{\linespread{1.2}
\begin{thrm}
\label{th.emb1b}
For any field configuration $u_i$,
if $\rho_a$ satisfies
(\ref{eq.elaux}) and $v_i$ is a solution of the linearization of (\ref{eq.de}) around $u_i$, then
$D_\mu j^\mu = 0$, where $j^\mu$ is given by (\ref{eq.elc2}) with $\delta u_i$ replaced by $v_i$.
The scaling transformation $\delta_{\mathrm{sc}} v_i = v_i$, $\delta_{\mathrm{sc}} \rho_a = -\rho_a$ is a symmetry of
the auxiliary Lagrangian associated with the linearization of (\ref{eq.de}) with $K^\mu=0$, and  
the corresponding Noether current is $j^\mu$. 
The relevant auxiliary system of DEs for $\rho_a$ coincides with (\ref{eq.elaux}).
\end{thrm} }
{\it Proof.}
Let $G^a$ denote the linearization of $F^a$. $G^a=0$ is a homogeneous linear DE system for the linearized fields
(being denoted by $v_i$),
therefore $\delta_{\mathrm{sc}} G^a = G^a$. 
The variation of the auxiliary Lagrangian under $\delta_{\mathrm{sc}}$ is thus
$\delta_{\mathrm{sc}}(G^a\rho_a) = \delta_{\mathrm{sc}} G^a \rho_a + G^a \delta_{\mathrm{sc}} \rho_a = 0$, 
i.e.\ $\delta_{\mathrm{sc}}$ is a symmetry of the auxiliary Lagrangian with $K^\mu=0$.
The statements that the Noether current is given by (\ref{eq.elc2}) and 
that the auxiliary system of DEs for $\rho_a$, i.e.\ the Euler--Lagrange equations for $v_i$,
coincides with (\ref{eq.elaux}) are not difficult to verify 
(in (\ref{eq.elc2}) the replacement $\delta u_i \to v_i$ is understood).
Note that $\delta_{\mathrm{sc}} \rho_a$ could be left arbitrary as well,
and then generally $\delta_{\mathrm{sc}}$ would only be an on-shell symmetry. $\Box$
 
It is clear that the current conservation laws stated in Theorem \ref{th.emb1} follow from Theorem \ref{th.emb1b}.
On the other hand, Theorem \ref{th.emb1} is sufficient to derive the conservation law stated in Theorem \ref{th.emb1b}.
It should be noted that $u_i$ is not required to be a solution of (\ref{eq.de}) in Theorem \ref{th.emb1b}.

Although the standard way of constructing conserved currents in the framework of the embedding method is 
described by Theorem \ref{th.emb1}, 
the Euler--Lagrange equation system consisting of (\ref{eq.de}) and (\ref{eq.elaux}) allows other 
possibilities. In particular, it allows the multiplier method to be connected with Noether's theorem:
{\linespread{1.2}
\begin{thrm}
\label{th.emb2}
$q_a[u]$ is a conservation law multiplier for $F^a[u]$ and for a current $J^\mu[u]$ if and only if
the one-parameter infinitesimal transformation of $\{ u_i, \rho_a \}$ that has the characteristic
$\delta_q u_i = 0$, $\delta_q\rho_a[u,\rho] = q_a[u]$ is a symmetry of $\hat{L}[u,\rho]$ with 
$K^\mu[u,\rho] = J^\mu[u]$. 
The Noether current corresponding to $\delta_q$ is $-J^\mu$.
\end{thrm} }
{\it Proof.} First, let us assume that $q_a$ is a conservation law multiplier for $F^a$ and $J^\mu$.  
Clearly $\delta_q F^a=0$, since $\delta_q u_i = 0$,
therefore $\delta_q\hat{L} = F^a\delta_q \rho_a+\delta_q F^a \rho_a = F^a q_a = D_\mu J^\mu$. This shows that 
$\delta_q$ is a symmetry of $\hat{L}$ with $K^\mu = J^\mu$. If $\delta_q$ is a symmetry of $\hat{L}$
with $K^\mu = J^\mu$,
then again $\delta_q\hat{L} = F^a\delta_q \rho_a+\delta_q F^a \rho_a = F^a q_a = D_\mu J^\mu$, 
i.e.\ $q_a$ is a conservation law multiplier for $F^a$ and $J^\mu$.
Since $\hat{L}$ does not depend on the derivatives of $\rho_a$ and $\delta_q u_i = 0$, the relevant $j^\mu$ current is zero,
therefore the Noether current corresponding to $\delta_q$ is $-K^\mu=-J^\mu$. $\Box$

Theorem \ref{th.emb2}, which cannot be found in the literature to our knowledge,
shows that from the point of view of the auxiliary Lagrangian (\ref{eq.l}), 
the multiplier method is a special case of Noether's theorem. 

It is important to note 
that constructing local conserved currents for (\ref{eq.de})
is not the only use of the embedding method, as can be seen from the second part of Theorem \ref{th.emb1}.
A standard application of conserved currents is the verification of numerical solutions of DEs, 
and for that
purpose $j^\mu$ can be constructed, without knowing any adjoint-symmetries, in the following way: 
for a given numerical solution of (\ref{eq.de}) that is to be verified
one computes a particular $\rho_a$ by numerically solving (\ref{eq.elaux}), 
and constructs $j^\mu$ using this $\rho_a$, the numerical solution of (\ref{eq.de}),
and a symmetry of (\ref{eq.de}). This $j^\mu$ should then be conserved with satisfactory precision.
See also \cite{t,t2} concerning this kind of application of the embedding method.

\section{Application of the two methods to extended systems of differential equations}
\label{sec.rel}

The aim of this section is to discuss our 
results about the embedding and the multiplier methods applied to 
the extended systems of DEs mentioned in the Introduction.
Before considering the extended systems, we rederive in Section \ref{sec.rel1} in slightly different form 
a fundamental result of \cite{Anco4, Anco1} (see also \cite{Anco3, AK}) 
on the connection between the two methods, according to which the current 
that the embedding method gives for a multiplier/symmetry pair is equivalent with the current 
obtained by the action of the symmetry on the current that belongs to the multiplier. 
The details of the derivation are included mainly because 
they differ from those in \cite{Anco4, Anco1}.
The special case when the current that belongs to the multiplier is homogeneous under the action of the symmetry,
and the relevance of this case for the question whether the currents that can be generated by the multiplier method can be 
reproduced by the embedding method as well, is also discussed.
Then in Section \ref{sec.ext} we introduce the extension of DEs in general form
and derive a variant of the result of Section \ref{sec.rel1} for them.
This includes a characterization of the multipliers of the extended DE systems.
The case of homogeneous currents is again considered.
A special class of adjoint-symmetries of the extended DEs is noted as well.
In Section \ref{sec.f} a particular way of extending an arbitrary system of DEs is discussed.
Finally examples of the application of the results of Sections \ref{sec.rel1} and \ref{sec.ext}
are presented in Section \ref{sec.examples}.

\subsection{Action of infinitesimal symmetries on conserved currents that correspond to a multiplier}
\label{sec.rel1}

If $q_a$ is a conservation law multiplier for $F^a$, then 
$\frac{\delta (F^a q_a)}{\delta u_i}=0$, and from this it follows that 
$\rho_a=q_a$ satisfies (\ref{eq.elaux}) if $F^a=0$. This means that if 
$q_a$ is a conservation law multiplier, then it is also an adjoint-symmetry.
In particular, if $F^a$ is the Euler--Lagrange derivative of a Lagrangian $L[u]$, 
then any $\delta u_i$ that determines a symmetry of $L$ is suitable for $\rho_a$ 
(the indices $i$ and $a$ have the same range 
if $F^a$ is an Euler--Lagrange derivative). It should be noted that in general it is not true that any adjoint-symmetry is a
conservation law multiplier \cite{Anco2}.

The variation of $F^a q_a$ under a one-parameter transformation $u_i \to u_i+\varsigma\,\delta u_i$ 
of $u_i$ is
\beq
\label{eq.dfq1}
\delta (F^a q_a) = 
\delta F^a q_a + F^a\delta q_a =
D_\mu (\delta J^\mu) =  \frac{\delta (F^a q_a)}{\delta u_i}\delta u_i + D_\mu j^\mu_{(Fq)}\, ,
\eeq
where $j^\mu_{(Fq)}$ denotes the current obtained by applying (\ref{eq.g2}) to $\delta (F^a q_a)$, 
with $\delta u_i$ in the role of $\epsilon^\alpha$,
and $\delta J^\mu$ is the first order variation of $J^\mu$ under the action of $u_i \to u_i+\varsigma\,\delta u_i$.
$j^\mu_{(Fq)}$ is given by the expression on the right hand side of (\ref{eq.elc2}), with $\rho_a$ replaced by $q_a$ 
($q_a$ depends on $u_i$ and its derivatives, 
therefore it is important to note that the replacement is understood to be done 
before any evaluation of the derivatives in (\ref{eq.elc2})).
As was noted above, 
(\ref{eq.mult}) implies that $\frac{\delta (F^a q_a)}{\delta u_i} = 0$, therefore from (\ref{eq.dfq1})
it follows that
\beq
\label{eq.vr}
D_\mu (j^\mu_{(Fq)} -\delta J^\mu)=0
\eeq
for arbitrary $u_i$, i.e.\ $j^\mu_{(Fq)} -\delta J^\mu$ is an identically conserved current.
Furthermore, the left part of (\ref{eq.dfq1}) shows that if $u_i \to u_i+\varsigma\,\delta u_i$ 
is a symmetry of (\ref{eq.de}), 
then $\delta J^\mu$ is also a conserved current. 
In virtue of (\ref{eq.vr}), $j^\mu_{(Fq)}$ is also conserved in this case and 
is equivalent with $\delta J^\mu$.

$j^\mu_{(Fq)}$ can be written as the sum 
\beq
\label{eq.jfq}
j^\mu_{(Fq)}=j^\mu_{(\tilde{F}q)}|_{\tilde{F}^a=F^a} + j^\mu_{(F\tilde{q})}|_{\tilde{q}_a=q_a}\, ,
\eeq
where $j^\mu_{(\tilde{F}q)}|_{\tilde{F}^a=F^a}$ is understood to be given by (\ref{eq.elc2}) in the following way:
$\rho_a$ should be replaced by $q_a$ and $F^a$ by $\tilde{F}^a$, 
where $\tilde{F}^a$ may depend on $x^\mu$ but not on $u_i$ and its derivatives, then the derivatives 
$\frac{\partial (\tilde{F}^a q_a)}{\partial (\partial_{\mu\dots}u_i)}$
should be evaluated,
and finally $\tilde{F}^a$ should be replaced again by $F^a$. 
$j^\mu_{(F\tilde{q})}|_{\tilde{q}_a=q_a}$ is understood in a similar way.
Up to possible signs depending on the commutation properties of the quantities involved, 
$\frac{\partial (\tilde{F}^a q_a)}{\partial (\partial_{\mu\dots}u_i)}=\tilde{F}^a\frac{\partial q_a}{\partial (\partial_{\mu\dots}u_i)}$ and 
$\frac{\partial (F^a \tilde{q}_a)}{\partial (\partial_{\mu\dots}u_i)}=\tilde{q}_a\frac{\partial F^a}{\partial (\partial_{\mu\dots}u_i)}$.
In different words, the two terms on the right hand side of (\ref{eq.jfq}) arise simply by 
applying the differentiation rule of products to the derivatives $\frac{\partial (F^a q_a)}{\partial (\partial_{\mu\dots}u_i)}$
in (\ref{eq.elc2}) (after $\rho_a$ has been replaced by $q_a$).

$j^\mu_{(F\tilde{q})}|_{\tilde{q}_a=q_a}$ is identical with $j^\mu$ given by (\ref{eq.elc2}), with $\rho_a=q_a$,
and if $F^a=0$, then obviously $j^\mu_{(\tilde{F}q)}|_{\tilde{F}^a=F^a}=0$, therefore 
$j^\mu$, with $\rho_a=q_a$, is equivalent with $\delta J^\mu$. 

The conclusions of the arguments above can be summarized in the following theorem, 
which connects the multiplier method with the embedding method:
\begin{thrm}
\label{th.1}
Let $u_i \to u_i+\varsigma\,\delta u_i$ be a symmetry of the system of differential equations (\ref{eq.de}), and
let $q_a$ be a conservation law multiplier for $F^a$ and $J^\mu$, i.e.
\beq
F^a[u] q_a[u] = D_\mu J^\mu[u]\, .
\eeq
Under these conditions $\delta J^\mu$ is also a conserved current, 
and the Noether current $j^\mu = j^\mu_{(F\rho)}|_{\rho_a=q_a}$
associated with the auxiliary Lagrangian $\hat{L}$, with the symmetry $u_i \to u_i+\varsigma\,\delta u_i$, 
and with the values $\rho_a=q_a$ of the auxiliary fields,
is equivalent with $\delta J^\mu$.
$j^\mu =  j^\mu_{(Fq)} -  j^\mu_{(\tilde{F}q)}|_{\tilde{F}^a=F^a}$ for arbitrary $u_i$, 
$j^\mu_{(\tilde{F}q)}|_{\tilde{F}^a=F^a}= 0$ on the solutions of (\ref{eq.de}),
thus $j^\mu$ is equivalent with $j^\mu_{(Fq)}$.
$j^\mu - \delta J^\mu + j^\mu_{(\tilde{F}q)}|_{\tilde{F}^a=F^a} =
j^\mu_{(Fq)} - \delta J^\mu$ is an identically conserved current.
\end{thrm}
Clearly $j^\mu$ is the current given by the embedding method for the symmetry characteristic $\delta u_i$ and for the 
adjoint-symmetry (and multiplier) $q_a$.

Theorem \ref{th.1} is slightly more general than Theorem 4 of \cite{Anco1}, 
as it only assumes about $F^a$ that it is differentiable as many times as necessary.
On the other hand, under the (mild) regularity assumptions on $F^a$
made in \cite{Anco1} (see also \cite{Anco2} for the definition of this regularity property) it is also possible 
to find a multiplier for $j^\mu$ explicitly, and this is actually done in \cite{Anco1}.
Nevertheless, for the purpose of the present paper it is not necessary to have a multiplier for $j^\mu$.
The proof of Theorem 4 in \cite{Anco1} is also different from the above derivation of Theorem \ref{th.1}, 
as it relies on the multiplier for $j^\mu$.

From (\ref{eq.dfq1}) and from the subsequent arguments it is clear that there is 
a variant of Theorem \ref{th.1} in which the condition
that $u_i \to u_i+\varsigma\,\delta u_i$ is a symmetry of (\ref{eq.de}) is replaced by the condition 
that $\delta(F^a q_a)=0$ if $F^a=0$:
\begin{prop}
\label{pr1.1}
Let $q_a$ be a conservation law multiplier for $F^a$ and $J^\mu$ and let 
$u_i \to u_i+\varsigma\,\delta u_i$ be a one-parameter transformation under the action of which
$\delta(F^a q_a)=0$ if $F^a=0$. Then 
$\delta J^\mu$, which is equivalent with $j^\mu = j^\mu_{(F\rho)}|_{\rho_a=q_a}$
and with $j^\mu_{(Fq)}$,
is also a conserved current.
$j^\mu_{(Fq)} - \delta J^\mu$ is identically conserved and 
$j^\mu_{(Fq)} - j^\mu =  j^\mu_{(\tilde{F}q)}|_{\tilde{F}^a=F^a}$ is zero on the solutions of (\ref{eq.de}). 
\end{prop}

\subsubsection{Homogeneous currents}
\label{sec.hc1}

Regarding the question whether the conserved currents that can be obtained 
by the multiplier method can also be obtained
by the embedding method, it follows from Theorem \ref{th.1} that 
if there is a symmetry $u_i \to u_i+\varsigma\,\delta u_i$ of (\ref{eq.de}) 
so that $\delta J^\mu$ is equivalent with $\omega J^\mu$, where $\omega$ is a nonzero constant, 
for a given current $J^\mu$ corresponding to a multiplier $q_a$,
then one can say that this $J^\mu$ can also be obtained by the embedding method up to equivalence, 
using $q_a$ as adjoint-symmetry and
$\delta u_i$ as symmetry characteristic in the construction.
This observation also has another version:
\begin{prop}
\label{pr1.2}
Let $q_a$ be a multiplier for $F^a$ and $J^\mu$,
let $u_i \to u_i+\varsigma\,\delta u_i$ be a one-parameter transformation of the fields,
and let $\omega$ be a constant. The following three statements are equivalent: 
{\rm (i)} $\delta(F^a q_a)=\omega F^a q_a$, {\rm (ii)}  
$\delta J^\mu$ differs from $\omega J^\mu$ by an
identically conserved current, 
{\rm (iii)} $j^\mu_{(Fq)}$ differs from $\omega J^\mu$ by an identically conserved current.
Since $j^\mu = j^\mu_{(F\rho)}|_{\rho_a=q_a}$ differs from $j^\mu_{(Fq)}$ by a current that is zero on the solutions of
(\ref{eq.de}), $j^\mu$ is also equivalent with $\omega J^\mu$ if 
{\rm (i)}, {\rm (ii)} or {\rm (iii)} holds. 
\end{prop}
(ii) follows from (i) in virtue of the equations
$D_\mu (\delta J^\mu) = \delta(F^a q_a) = \omega F^a q_a =\omega D_\mu J^\mu$.
If (ii) holds, then $0=D_\mu (\delta J^\mu - \omega J^\mu)=\delta(F^a q_a) - \omega F^a q_a$, thus 
(i) also holds.
(ii) and (iii) are equivalent because $j^\mu_{(Fq)} - \delta J^\mu$ is identically conserved.

In particular, if $\delta F^a=\omega_F F^a$ and $\delta q_a=\omega_q q_a$ with some constants $\omega_F^a$
and $\omega_q$, 
then $\delta(F^a q_a)=\omega  F^a q_a$ with $\omega=\omega_F+\omega_q$, and if 
$\omega_F+\omega_q\ne 0$, then again one can say that   
$J^\mu$ can also be obtained by the embedding method up to equivalence, 
using $q_a$ as adjoint-symmetry and
$\delta u_i$ as symmetry characteristic. Note that the conditions 
$\delta F^a=\omega_F F^a$, $\delta q_a = \omega_q q_a$, $\omega\ne 0$ can be verified without knowing $J^\mu$,
and $\delta F^a=\omega_F F^a$ implies that $u_i \to u_i+\varsigma\,\delta u_i$ is a symmetry of (\ref{eq.de}).

Usual symmetries under which the equivalence of $\delta J^\mu$ with $\omega J^\mu$
(or even $\delta J^\mu = \omega J^\mu$),
or $\delta F^\mu = \omega_F F^\mu$ and $\delta q_a = \omega_q q_a$
can occur are the scaling symmetries, and for this reason
they will have an important role in Sections \ref{sec.f} and \ref{sec.examples}.

Note that Proposition \ref{pr1.2} allows one to calculate $J^\mu$ up to equivalence if $q_a$ is known, (i) holds
and $\omega\ne 0$. 
This is a variant of one of the methods mentioned in \cite{Anco2} (see also \cite{Anco4, Anco3, AK}) 
for calculating the current that corresponds to a given multiplier.

\subsection{The extended systems of differential equations}
\label{sec.ext}

Let us assume that the $F^a$ that defines the system (\ref{eq.de}) also contains some constant parameters, 
which will be denoted by $g_l$. These $g_l$ may be only a subset of all constant parameters in $F^a$.
If $F^a$ describes a physical system, 
then $g_l$ can be coupling constants and dimensionful scale parameters, for example.
It may be useful to introduce some new parameters into the original system of DEs as well.
In accordance with the presence of the parameters in the differential equations, 
the conserved currents may also depend on the parameters, and it will be assumed that they are conserved for 
arbitrary values of the parameters. Although there may be currents that are conserved only for special values of
the parameters, they will not be considered in this paper. 

$g_l$ can be regarded as additional fields satisfying the DEs $\partial_\mu g_l = 0$,
i.e.\ one can consider the \emph{extended system of DEs}
\beq
\label{eq.deext}
F^a[u,g]=0\, ,\qquad \partial_\mu g_l=0
\eeq
instead of (\ref{eq.de}). 
As $g_l$ are originally constant parameters,
$F^a[u,g]$ does not depend on the derivatives of $g_l$.
In the following we derive a variant of Theorem \ref{th.1}
for the DE systems (\ref{eq.deext}) and give a characterization of their multipliers.

The auxiliary Lagrangian density function for the extended system (\ref{eq.deext}) is
\beq
\label{eq.cL}
\check{L}[u,g,\rho,\vartheta] = F^a[u,g]\rho_a + \partial_\mu g_l\, \vartheta^{\mu l}\, , 
\eeq
where $\vartheta^{\mu l}$ are further auxiliary fields corresponding to the equations $\partial_\mu g_l = 0$. 
The Euler--Lagrange derivatives 
of $\check{L}$ with respect to the fields are
\beq
\label{eq.ext1}
\frac{\delta \check{L}}{\delta \rho_a} = F^a\, ,
\qquad \frac{\delta \check{L}}{\delta \vartheta^{\mu l}} = \partial_\mu g_l\, ,
\eeq
\beq
\label{eq.ext2}
\frac{\delta \check{L}}{\delta u_i} = \frac{\delta (F^a \rho_a)}{\delta u_i}\, ,
\qquad \frac{\delta \check{L}}{\delta g_l} = \frac{\partial F^a}{\partial g_l} \rho_a - \partial_\mu \vartheta^{\mu l}\, .
\eeq
From (\ref{eq.ext1}) and (\ref{eq.ext2}) it can be seen that the equations for $\rho_a$ 
are not changed by appending $\partial_\mu g_l = 0$,
and the only Euler--Lagrange equations that contain $\vartheta^{\mu l}$ 
are $\frac{\partial F^a}{\partial g_l} \rho_a - \partial_\mu \vartheta^{\mu l}=0$.

Let us assume that a one-parameter transformation specified by 
a characteristic $\{ \delta u_i, \delta g_l\}$ is 
a symmetry of the extended system (\ref{eq.deext}).
The associated conserved current, given by the second part of Theorem \ref{th.emb1}
applied to (\ref{eq.deext}), is
\beq
\label{eq.cons}
j^\mu = j^\mu_{(F\rho)} + \vartheta^{\mu l} \delta g_l\, , 
\eeq
where $j^\mu_{(F\rho)}$ denotes the expression (\ref{eq.elc2}). 
Although $\vartheta^{\mu l}$ is not known, it can, in principle, 
be obtained by solving the Euler--Lagrange equations
$\frac{\partial F^a}{\partial g_l} \rho_a - \partial_\mu \vartheta^{\mu l}=0$.
Note that $\delta g_l$ are constants for the solutions of (\ref{eq.deext}).
Clearly if $\vartheta^{\mu l}[u,g]$ is a conserved current of (\ref{eq.deext}) for all values of $l$, 
then $\{ \rho_a=0, \vartheta^{\mu l}[u,g]\}$
is an adjoint-symmetry of (\ref{eq.deext}), $j^\mu_{(F\rho)}=0$, and thus $j^\mu = \vartheta^{\mu l} \delta g_l$.

$q_a[u,g]$ will be called a \emph{parameterized multiplier} for $F^a$, if (a)
$q_a[u,g]$ is a conservation law multiplier for $F^a$ in the sense that
\beq
\label{eq.mg}
F^a[u,g]q_a[u,g]=D_\mu J^\mu[u,g]
\eeq
holds with some $J^\mu[u,g]$ for arbitrary $u_i$ and arbitrary but constant $g_l$,
(b) $q_a[u,g]$ and $J^\mu[u,g]$, which are local functions of $g_l$,
do not depend on the derivatives of $g_l$.
A parameterized multiplier is a multiplier in the usual sense, but it can depend on the parameters of  
$F^a$ and it is required to be a multiplier for arbitrary values of these parameters. 
$J^\mu[u,g]$ will be called a \emph{parameterized conserved current}.

From (\ref{eq.mg}) it is easy to see that if $q_a[u,g]$ is a parameterized multiplier, then
\beq
\label{eq.pm}
F^a[u,g]q_a[u,g] + \frac{\partial J^\mu[u,g]}{\partial g_l}\partial_\mu g_l = D_\mu J^\mu[u,g]
\eeq
holds for general (i.e.\ not necessarily constant) $g_l$, therefore $\{ q_a, \frac{\partial J^\mu}{\partial g_l} \}$ is a (ordinary) conservation law multiplier 
for the extended system (\ref{eq.deext}).

The considerations above together with Theorem \ref{th.1} give the following variant of Theorem \ref{th.1}
for extended DE systems:
\begin{thrm}
\label{th.2}
Let a one-parameter transformation with characteristic
$\{\delta u_i, \delta g_l\}$ 
be a symmetry of the extended system of differential equations
\beq 
\label{eq.exd}
F^a[u,g]=0\, , \qquad \partial_\mu g_l=0
\eeq
introduced above. 
Let $q_a[u,g]$ be a parameterized conservation law multiplier for $F^a[u,g]$ 
and for the current $J^\mu[u,g]$. 
Then $\{ q_a, \frac{\partial J^\mu}{\partial g_l} \}$ is a conservation law multiplier for (\ref{eq.exd}) 
with the same current $J^\mu[u,g]$.
The Noether current (\ref{eq.cons}) associated with
the symmetry specified by $\{ \delta u_i, \delta g_l\}$,
with the auxiliary Lagrangian $\check{L}$, 
and with the values $\rho_a=q_a$ and $\vartheta^{\mu l}=\frac{\partial J^\mu}{\partial g_l}$ of the auxiliary fields, is
\beq 
\label{eq.jth}
j^\mu = j^\mu_{(F\rho)}|_{\rho_a=q_a} + \frac{\partial J^\mu}{\partial g_l} \delta g_l\, .
\eeq
$j^\mu$ is equivalent with $\delta J^\mu$, and if $g_l$ are regarded as constants, then
$j^\mu - \delta J^\mu + j^\mu_{(\tilde{F}q)}|_{\tilde{F}^a=F^a}$ is an identically conserved current.
\end{thrm}
Similarly as in Theorem \ref{th.1},
$j^\mu$ is the current given by the embedding method for the symmetry characteristic $\{ \delta u_i, \delta g_l\}$
and for the adjoint-symmetry (and multiplier) $\{ q_a, \frac{\partial J^\mu}{\partial g_l} \}$.

Since Theorem \ref{th.2} appears to deal with a special class of the multipliers and conserved currents of (\ref{eq.deext}),
one can ask whether 
there can be other, significantly different multipliers and corresponding conserved currents. 
Regarding this question, it is not difficult to verify the following:
\begin{prop}
\label{pr.1}
Let $\{ \tilde{q}_a[u,g], \theta^{\mu l}[u,g]\}$ be a multiplier for (\ref{eq.deext}) and for the current $\tilde{J}^\mu[u,g]$ 
(i.e.\ $F^a \tilde{q}_a + \partial_\mu g_l\, \theta^{\mu l} = D_\mu \tilde{J}^{\mu}$).
Then 
$q_a(x,u,\partial u,\dots,g)=
\tilde{q}_a(x,u,\partial u,\dots, g, 0, \dots)$
is a parameterized conservation\hspace{0.5mm} law\hspace{0.5mm} multiplier\hspace{0.5mm} for\hspace{0.5mm} $F^a[u,g]$\hspace{0.5mm} and\hspace{0.5mm} for\hspace{0.5mm} the\hspace{0.5mm} current\hspace{0.5mm}
$J^\mu(x,  u,  \partial u,\dots,g)
=\tilde{J}^\mu(x,  u, \partial u, \dots, g, 0,\\ \dots)$.
\end{prop}
Since $J^\mu=\tilde{J}^\mu$ if $\partial_\mu g_l=0$, Proposition \ref{pr.1} shows that   
the multipliers of (\ref{eq.deext}) considered in Theorem \ref{th.2} are sufficiently general.

When one searches for conserved currents using the multiplier method, the first step is to find multipliers, and
at this stage the conserved currents are not known. It is thus interesting to ask whether
the $\frac{\partial J^\mu}{\partial g_l}$ part of the multipliers of 
(\ref{eq.deext}) appearing in Theorem \ref{th.2} is unique. The following answer can be given to this question:
\begin{prop}
\label{pr.2}
Let both $\{ q_a[u,g], \theta_1^{\mu l}[u,g] \}$ and $\{ q_a[u,g], \theta_2^{\mu l}[u,g] \}$,
where $q_a[u,g]$, $\theta_1^{\mu l}[u,g]$ and $\theta_2^{\mu l}[u,g]$ do not depend on the derivatives of $g_l$, 
be multipliers for (\ref{eq.deext}) and for the currents $J_1^\mu[u,g]$ and $J_2^\mu[u,g]$ that differ only by an identically
conserved current. Then $\theta_1^{\mu l}=\theta_2^{\mu l}$. 
\end{prop}
{\it Proof.} From the conditions it follows that 
$(\theta_1^{\mu l} - \theta_2^{\mu l})\partial_\mu g_l =0$ holds for any configuration of 
$u_i$ and $g_l$. This implies $\theta_1^{\mu l} - \theta_2^{\mu l}=0$, because 
$\theta_1^{\mu l} - \theta_2^{\mu l}$ does not depend on the derivatives of $g_l$
and at any $x^\mu$ $\partial_\mu g_l$ can be chosen arbitrarily and independently of 
$u_i(x)$ and $g_l(x)$. $\Box$

From Theorem \ref{th.2} and Propositions \ref{pr.1} and \ref{pr.2} the following conclusion can be drawn:
{\linespread{1.2}
\begin{thrm}
Let $\{ q_a[u,g], \theta^{\mu l}[u,g] \}$, where $q_a[u,g]$ and $\theta^{\mu l}[u,g]$ do not depend on
the derivatives of
$g_l$, be a multiplier for (\ref{eq.deext}) and for a current $\tilde{J}^\mu[u,g]$.
Then $q_a[u,g]$ is a parameterized multiplier
for $F^a[u,g]$ and for the current
$J^\mu (x,  u,  \partial u,\dots,g)
=\tilde{J}^\mu (x,  u, \partial u, \dots,g, 0,\dots)$,
$\{ q_a[u,g], \theta^{\mu l}[u,g] \}$ is a multiplier
for $J^\mu [u,g]$, and $\theta^{\mu l}=\frac{\partial J^\mu}{\partial g_l}$.
\end{thrm} }
This implies, together with Theorem \ref{th.2} and Proposition \ref{pr.1}, 
that when one looks for multipliers of (\ref{eq.deext}) it is sufficient to consider only the multipliers that
do not depend on the derivatives of $g_l$,
and these multipliers are the same as those that appear in Theorem \ref{th.2} (i.e.\ $\{ q_a, \frac{\partial J^\mu}{\partial g_l} \}$).
The currents that correspond to other multipliers are equivalent with those that correspond to the
multipliers appearing in Theorem \ref{th.2}.

The fact that the multipliers $\{ q_a, \frac{\partial J^\mu}{\partial g_l} \}$ of (\ref{eq.deext}) contain
$\frac{\partial J^\mu}{\partial g_l}$ is useful for solving the problem of finding the conserved current corresponding to
a known multiplier, since if one has found $J^\mu$ for a special value of $g_l$,
then it can be calculated for other values of $g_l$ by integrating $\frac{\partial J^\mu}{\partial g_l}$
with respect to $g_l$.

\subsubsection{Homogeneous currents}

Turning again to the question whether the conserved currents that can be obtained
by the conservation law multiplier method can also be obtained
by the embedding method,
it is generally not obvious to what extent the symmetries of (\ref{eq.de}) allow the conserved currents 
to be reproduced in the way mentioned in Section \ref{sec.hc1}, i.e.\ 
whether for a given conserved current there is a symmetry under the action of which it is homogeneous
with nonzero weight,
but in Sections \ref{sec.f} and \ref{sec.examples} it will be demonstrated that
considering extended systems (\ref{eq.deext}) can be useful in this respect,
as the extended systems can have the necessary symmetries.
The main idea that will be followed is to try
to extend (\ref{eq.de}) in such a way 
that the extended DE system has a simple scaling symmetry under which the 
conserved currents of interest are homogeneous with nonzero scaling weight.
This idea appears also in \cite{Anco2} in the context of the problem of calculating the 
conserved current that corresponds to a given multiplier. The explanation for the relevance of
the same idea for both the latter problem and for the problem considered in this paper, 
i.e.\ for the problem of reproducing conserved currents that correspond to
a multiplier by the embedding method,
is given by Section \ref{sec.hc1}. 

Under the conditions of Theorem \ref{th.2}, if $\delta J^\mu$ is equivalent with $\omega J^\mu$, 
where $\omega$ is a nonzero constant, 
then it can be said that $J^\mu$ can be reproduced by the embedding method. Nevertheless, from the point of view
of the original DE system it is satisfactory if $\delta J^\mu$ is equivalent with $\omega J^\mu$ 
only for constant $g_l$, 
by which we mean that $\delta J^\mu[u,g]-\omega J^\mu[u,g] = \bar{J}^\mu[u,g]+\hat{J}^\mu[u,g]$, where
$\bar{J}^\mu[u,g]$ is conserved for arbitrary $u_i$ and constant $g_l$, and $\hat{J}^\mu[u,g]=0$ if
$g_l$ is constant and $F^a=0$. 
For deciding whether this equivalence holds, the following observation can be useful:
\begin{prop}
\label{pr.scc}
Let $q_a[u,g]$ be a parameterized multiplier for $F^a[u,g]$ and $J^\mu[u,g]$, $\omega$ a constant, and
$\{ \delta u_i,\delta g_l\}$ the characteristic of a  one-parameter transformation of $\{ u_i , g_l \}$.
If {\rm (i)} $\delta(F^a q_a)= \omega F^a q_a$ for arbitrary $u_i$ and $g_l$ and 
{\rm (ii)} $D_\mu \delta g_l=0$ for constant $g_l$, 
then $\delta J^\mu -\omega J^\mu$
is conserved for arbitrary $u_i$ and constant $g_l$. 
\end{prop}
Note that the conditions $\delta(F^a q_a)= \omega F^a q_a$ and $D_\mu \delta g_l=0$
can be verified even if $J^\mu$ is not known.
To derive Proposition \ref{pr.scc}, let us consider the variation of (\ref{eq.pm}): 
$\delta(F^a q_a) +\delta (\frac{\partial J^\mu}{\partial g_l})\partial_\mu g_l + 
\frac{\partial J^\mu}{\partial g_l} D_\mu \delta g_l = D_\mu \delta J^\mu$.
From this equation it follows that if the conditions of the proposition hold, 
then $D_\mu (\omega J^\mu) = \omega F^a q_a = \delta(F^a q_a) = D_\mu \delta J^\mu$ for constant $g_l$
and arbitrary $u_i$.

\subsubsection{Using adjoint-symmetries of the form $\{0, \vartheta^{\mu l} \} $}
\label{sec.as}

There are also other possibilities, distinct from those mentioned above,
for reproducing the conserved currents of (\ref{eq.deext})
by the embedding method, 
based on the observation below (\ref{eq.cons}) about the adjoint-symmetries of (\ref{eq.deext}). In particular,
the following holds for any conserved current of (\ref{eq.deext}):
\begin{prop}
\label{pr.sg2}
Let $\{ \delta u_i, \delta g_l \}$ be the characteristic of a symmetry of (\ref{eq.deext}), 
let $\vartheta^\mu[u,g]$ 
be a conserved current of (\ref{eq.deext}), and let $l_0$ be a value in the range of the index $l$.
Then $\{0,\vartheta^\mu \delta^l_{l_0}\}$ (where $\delta^l_{l_0}$ is a Kronecker delta) is an adjoint-symmetry of (\ref{eq.deext}) and the conserved current produced by the embedding method 
using this adjoint-symmetry and the symmetry characteristic $\{ \delta u_i, \delta g_l \}$ is $j^\mu=\vartheta^\mu\delta g_{l_0}$.
\end{prop}
Since $\delta g_l$ is constant on the solutions of (\ref{eq.deext}), one can say that under the conditions of the proposition $\vartheta^\mu$ 
can be reproduced by the embedding method for all values of $g_l$ for which $\delta g_{l_0}\ne 0$. 
Note that the conditions of Proposition \ref{pr.sg2} are rather mild, in particular
$\vartheta^\mu$ is not required to correspond to a multiplier, and only $\delta g_{l_0}\ne 0$ is necessary 
for $j^\mu$ to be nontrivial. On the other hand, Proposition \ref{pr.sg2} does not have much practical use for 
constructing conserved currents,
since it requires the knowledge of $\vartheta^\mu$ in advance.

\subsection{A special extension of arbitrary systems of differential equations}
\label{sec.f}

In this section we study the following trivial extension of an arbitrary system of DEs $F^a[u] = 0$:
\beq
\label{eq.gf}
F^a[u] = 0\,,\qquad \partial_\mu g = 0\, ,
\eeq
where $g$ is a single real valued field.
This extension is trivial, as $F^a[u]$ does not actually depend on $g$. 
(\ref{eq.gf}) obviously has the scaling symmetry
\beq
\label{eq.scl}
\delta u_i=0\,, \qquad \delta g = g\, ,
\eeq 
which is also completely independent of $F^a[u]$. Although (\ref{eq.gf}) is not useful from a practical point of view,
it deserves consideration as the simplest extension of arbitrary DEs and it gives some further insight, beyond 
the results of Section \ref{sec.ext},
into the nature of the extended systems.

Concerning the embedding method, the following statements can be made:
\begin{prop}
\label{th.3}
{\rm (i)} $\{ \rho_a[u], J^\mu[u]\}$ is an adjoint-symmetry of (\ref{eq.gf}) if and only if
$J^\mu[u]$ is a conserved current and $\rho_a[u]$ is an adjoint-symmetry of 
the DE system $F^a[u]=0$.
The conserved current (\ref{eq.cons}) associated with $\{ \rho_a[u],J^\mu[u]\}$ and with 
the symmetry (\ref{eq.scl}) is $gJ^\mu[u]$.
{\rm (ii)} $\{ 0, J^\mu[u]\}$ is an adjoint-symmetry of (\ref{eq.gf}) for any conserved current $J^\mu[u]$ of 
$F^a[u]=0$.
\end{prop}
Proposition \ref{th.3}
implies that all local conserved currents of (\ref{eq.de}) can be obtained 
by applying the embedding method to (\ref{eq.gf}).

The first part follows from the Euler--Lagrange equations corresponding to (\ref{eq.ext2}),
which take the form $\frac{\delta(F^a\rho_a)}{\delta u_i}=0$,
$\partial_\mu \vartheta^\mu=0$ in the case of (\ref{eq.gf}), and from (\ref{eq.scl}). 
In particular, $\delta u_i = 0$ implies that $j^\mu_{(F\rho)}=0$,
thus only the second term remains on the right hand side of (\ref{eq.cons}).
The second part of the proposition is an obvious corollary of the first part. 

The next two statements concern multipliers: 
\begin{prop}
\label{th.4}
Let $\{ gq_a[u],\vartheta^\mu[u]\}$ be a multiplier for (\ref{eq.gf}) and for a conserved current $J^\mu[u,g]$.
Then {\rm (i)} $\vartheta^\mu[u]$ is a conserved current of $F^a[u]=0$ with the multiplier $q_a[u]$, 
{\rm (ii)} $g\vartheta^\mu[u]$ is a conserved current of (\ref{eq.gf}) with the multiplier 
$\{ gq_a[u],\vartheta^\mu[u]\}$,
and thus $g\vartheta^\mu[u]$ differs from $J^\mu[u,g]$ by an identically conserved current.
\end{prop}
The first part of Proposition \ref{th.4} can be proved by 
taking the Euler--Lagrange derivative of the multiplier identity
$gF^a[u]q_a[u] + \partial_\mu g \vartheta^\mu[u]  = D_\mu J^\mu[u,g]$ with respect to $g$.
This gives $F^a[u]q_a[u] - D_\mu  \vartheta^\mu[u] = 0$, which is the sought result.
For the second part, one calculates $D_\mu (g\vartheta^\mu)$, also using the first part:
$D_\mu (g\vartheta^\mu) = \vartheta^\mu \partial_\mu g + g D_\mu \vartheta^\mu=
\vartheta^\mu \partial_\mu g + gF^a q_a = D_\mu J^\mu$.
\begin{prop}
\label{th.5}
Let $q_a[u]$ be a conservation law multiplier for $F^a[u]=0$ and $J^\mu[u]$.
Then $\{ gq_a, J^\mu\}$ is a multiplier for the extended system (\ref{eq.gf}) 
and the corresponding conserved current is $gJ^\mu$.
\end{prop}
Proposition \ref{th.5} follows from the equations $D_\mu(gJ^\mu)=D_\mu J^\mu g + J^\mu\partial_\mu g
=F^a q_a g + J^\mu\partial_\mu g$.

Propositions \hfill \ref{th.4} \hfill and \hfill \ref{th.5} \hfill mean \hfill that \hfill for \hfill any \hfill pair \hfill 
$(q_a[u], J^\mu[u])$ \hfill consisting \hfill of \hfill a \hfill multiplier \hfill
and \hfill a \hfill
corresponding \hfill conserved \hfill current \hfill
of \hfill the \hfill original \hfill system \hfill of \hfill DEs, \hfill there \hfill is \hfill a \hfill corresponding \hfill pair\\
$(\{ gq_a[u], J^\mu[u]\}, gJ^\mu[u])$
for the extended system (\ref{eq.gf}), and the converse is also true.

$gq_a[u]$ is also a parameterized multiplier for $F^a[u]$ and $gJ^\mu[u]$, and $\delta(gJ^\mu[u]) = gJ^\mu[u]$,
therefore the current $j^\mu$ given by Theorem \ref{th.2} for $gq_a[u]$ and for the scaling symmetry (\ref{eq.scl}) is 
equivalent with $gJ^\mu[u]$. Moreover, it is easy to see that $j^\mu$ is in fact equal to $gJ^\mu$.
This shows that any local conserved current of (\ref{eq.de}) that corresponds to
a multiplier can be reproduced by applying the embedding method to (\ref{eq.gf}) in such a way that the adjoint-symmetry that 
is used is also a multiplier.

Propositions \ref{th.4} and \ref{th.5} also imply that by searching for the multipliers of (\ref{eq.gf}) that have the form
$\{ gq_a[u], J^\mu[u]\}$, one can find the multipliers of the original system
$F^a[u]=0$ together with the corresponding conserved currents in one step, 
and it is not necessary to solve separately the problem of finding the conserved currents that correspond to the multipliers 
of $F^a[u]=0$. 

Other essentially trivial extensions of (\ref{eq.de}) could also be considered; 
for example $gF^a=0$, $\partial_\mu g = 0$, with $g(x)\ne 0$ for any $x^\mu$. It is easy to see that  
the above results, with minor modifications, hold for the latter extended DEs as well.

In the next section we discuss examples of other possibilities of extending the original system of equations
and endowing it with a suitable scaling symmetry, which are more specific to the features of the original DEs
and are not trivial.

\subsection{Examples}
\label{sec.examples}

In this section the application of the results of Sections \ref{sec.rel1} and \ref{sec.ext} in four special cases is discussed.
In each case, the aim is to show how certain currents corresponding to some multipliers are reproduced by the embedding
method, making use of extended DE systems of the form (\ref{eq.deext}).
The first example is an exception as it does not require any actual extension,
nevertheless it is included because of its fundamental nature.
Although in the second, third and fourth examples the discussion begins, for concreteness,
with relatively special differential equations,
these special cases have straightforward generalizations to wide classes of DE systems, 
which are also discussed briefly in the last paragraphs of the relevant subsections.
The fourth example, the generalized Korteweg--de Vries equation, is chosen partly because
it allows us to rely on the results of \cite{Anco2,AK} about this equation.
The DEs in the first example are linear, whereas in the other examples they are sums of a linear and a nonlinear part.
The symmetry transformations needed in Theorems \ref{th.1} and \ref{th.2} will be scaling transformations,
and they will be indicated by the subscript ${}_{\mathrm{sc}}$.
Although we focus on the use of scaling symmetries in conjunction with Theorems \ref{th.1} and \ref{th.2},
Proposition \ref{pr.sg2} also applies to all examples except the first one.
In Sections \ref{sec.ex2} and \ref{sec.ex3}, spacetime indices are understood to be raised and lowered by the Minkowski metric.

\subsubsection{Homogeneous linear differential equations}

Let us assume that $F^a[u]$ is homogeneous linear in $u_i$, i.e.\ it is of the form
$R\indices{^{ai}}u_i+R\indices{^{a\mu i}}\partial_\mu u_i+ R\indices{^{a\mu\nu i}}\partial_{\mu\nu} u_i+\dots$,
where $R\indices{^{a\dots i}}(x)$ are coefficient functions.
In this case the variation of $F^a$ under the scaling transformation $\delta_{\mathrm{sc}}u_i = u_i$
is $\delta_{\mathrm{sc}} F^a=F^a$, thus the system $F^a=0$ has a scaling symmetry even in itself.
If, in addition, 
a current $J^\mu[u]$, which corresponds to some multiplier, 
is a homogeneous $n$-th order expression of $u_i$ and its derivatives ($n$ is $1$ or $2$ in most cases), 
then $\delta_{\mathrm{sc}} J^\mu = nJ^\mu$,
therefore $nJ^\mu$ is equivalent with the current $j^\mu$ appearing in Theorem \ref{th.1}.

\subsubsection{Klein--Gordon field with $\phi^n$ interaction term}
\label{sec.ex2}

Let us consider a relativistic scalar field $\phi$ with the Lagrangian
$L=\frac{1}{2}(\partial_\mu \phi \partial^\mu \phi - m^2 \phi^2) + g\phi^n$, where $g\in\RR$ is a coupling constant, 
$m\in\RR$ is a mass parameter and $n\in \RR\setminus\{0\}$.
The field equation for $\phi$ is $F=0$, where $F$ is the
Euler--Lagrange derivative of $L$ with respect to $\phi$, i.e.
\beq
F=\frac{\delta L}{\delta \phi}= -\partial_\mu \partial^\mu\phi - m^2\phi +gn\phi^{n-1}\, .
\eeq
It is well-known that spacetime translations are symmetries of $L$. 
The variation of $\phi$ and $L$ under a spacetime translation in the direction given by a vector $h^\mu$ 
is $\delta\phi=-h^\mu\partial_\mu \phi$ and
$\delta L = -D_\mu(h^\mu L)$. The latter equation shows that $K^\mu = -h^\mu L$ can be chosen in (\ref{eq.K}).
$j^\mu = \frac{\partial L}{\partial(\partial_\mu\phi)}\delta\phi = - h^\nu \partial^\mu\phi \partial_\nu\phi$ 
(see \ref{eq.elc}),
thus the Noether current associated with spacetime translations in the direction $h^\mu$ is 
\beq 
\label{eq.jn}
J_N^\mu = -(\partial^\mu\phi\, \partial_\nu \phi - \delta^\mu_\nu L)  h^\nu\, .
\eeq

For applying the constructions described in Section \ref{sec.ext}, we promote $g$ to a field, 
i.e.\ we consider the extended system 
of equations $F=0$, $\partial_\mu g=0$. $m$, on the other hand, is left as a constant parameter.
The variation of $F$ and $\partial_\mu g$
under the scaling transformation 
\beq 
\label{eq.scaling}
\delta_{\mathrm{sc}} \phi = \phi\, ,\qquad \delta_{\mathrm{sc}} g = (2-n)g
\eeq
is $\delta_{\mathrm{sc}} F = F$ and $\delta_{\mathrm{sc}} (\partial_\mu g) = (2-n) \partial_\mu g$,
thus this scaling transformation is a symmetry of the extended system. 
$J^\mu_N$ also transforms homogeneously under (\ref{eq.scaling}):
\beq
\delta_{\mathrm{sc}}J^\mu_N = 2J^\mu_N\, .
\eeq

In accordance with the remark at the end of Section \ref{sec.clm},
$q=\delta\phi=-h^\mu\partial_\mu \phi$ is a multiplier for $F$ and for the conserved current $J^\mu=-J^\mu_N$. 
Moreover, $q$ is also a parameterized conservation law multiplier, the parameter being $g$, with the same current $J^\mu$.
For $j^\mu_{(F\rho)}$ one finds $j_{(F\rho)}^\mu = -\rho\,\partial^\mu\phi + \partial^\mu \rho\, \phi$, thus
$j_{(F\rho)}^\mu |_{\rho=q} = (\partial^\mu\phi\, \partial_\lambda \phi - \phi\,\partial^\mu\partial_\lambda \phi)h^\lambda$.
$\frac{\partial J^\mu}{\partial g} = -h^\mu \phi^n$, and finally (\ref{eq.jth}) gives
\beq 
\label{eq.jres}
j^\mu = (\partial^\mu\phi\, \partial_\lambda \phi - \phi\,\partial^\mu\partial_\lambda \phi)h^\lambda  + g(n-2)h^\mu \phi^n\, .
\eeq
In contrast with $J^\mu$, this current depends on the second derivatives of $\phi$,
and the coefficient of $\phi^n$ is also different in the two currents.
For $j^\mu_{(\tilde{F} q)}|_{\tilde{F}=F}$ one finds $j^\mu_{(\tilde{F} q)}|_{\tilde{F}=F} = -Fh^\mu\phi$, 
and then it is not difficult to verify that 
\beq 
\label{eq.jjj}
2J^\mu- j^\mu - j^\mu_{(\tilde{F} q)}|_{\tilde{F}=F}  = \partial_\nu (h^\nu \phi\, \partial^\mu \phi - h^\mu \phi\, \partial^\nu\phi)\, .
\eeq
This is an identically conserved current, in accordance with Theorem \ref{th.2}, thus $j^\mu$ is equivalent with $2J^\mu$.
Since (\ref{eq.jjj}) does not depend on $g$, it is identically conserved even when $g$ is not constant.

In addition to translation symmetries $L$ has Lorentz symmetry as well, which could also be used in the embedding method 
instead of the scaling symmetry,
without any extension of the original Euler--Lagrange equation.
Nevertheless, the extended DE system is also suitable for dealing with more general cases that do not have Lorentz symmetry.
Consider, for instance, the Lagrangian 
$L=\frac{1}{2}(\partial_\mu \phi \partial^\mu \phi - m^2 \phi^2) + gW\phi^n$, where $W$ is an arbitrary function of all coordinates
except $x^0$. This Lagrangian still has $x^0$-translation symmetry, but generally does not have Lorentz symmetry.
The preceding steps can nevertheless be repeated essentially identically;
$F=-\partial_\mu \partial^\mu \phi -m^2\phi+gWn\phi^{n-1}$, 
$J_N^\mu = - \partial^\mu\phi \partial_0\phi - \delta_0^\mu L$,
$\delta_{\mathrm{sc}}J_N^\mu = 2J_N^\mu$,
$\frac{\partial J^\mu}{\partial g} = -\delta_0^\mu W\phi^n$,
and finally
$j^\mu = (\partial^\mu\phi\, \partial_0 \phi - \phi\,\partial^\mu\partial_0 \phi) + g(n-2)W \phi^n \delta_0^\mu$,
which is equivalent with $2J^\mu = -2J^\mu_{N}$.

Clearly the case when the interaction term is $\sum_k g_k \phi^{n_k}$ or $\sum_k g_k W_k \phi^{n_k}$ can be treated very similarly
to the cases above, promoting each coefficient $g_k$ to a field with transformation property $\delta_{\mathrm{sc}}g_k = (2-n_k)g_k$. 
Moreover, the same treatment would also be suitable for several fields with various tensorial structure 
and with a Lagrangian of the form 
\beq 
L = L_0[u]+\sum_k g_k L_k[u]\, ,
\eeq
where $L_0$ is a quadratic local function of the fields and
$L_k$ is of order $n_k$.
The symmetry of $L$ does not have to be a translation; it is sufficient that $\delta u_i$ and $K^\mu$ be homogeneous under 
$\delta_{\mathrm{sc}}$ of order $\omega\ne -1$ and $1+\omega$, respectively. Under these conditions $J^\mu_N$ is also 
homogeneous of order $1+\omega$. For the usual spacetime and internal symmetries $\omega=1$.

\subsubsection{Klein--Gordon field with general interaction}
\label{sec.ex3}

In the previous example we made use of the homogeneous polynomial form of the interaction term, 
thus it is natural to ask what can be done in the more general case when the Lagrangian is
$L=\frac{1}{2}(\partial_\mu \phi \partial^\mu \phi - m^2 \phi^2) + V(\phi)$, where $V(\phi)$ is an arbitrary function.
A relatively simple way to deal with this case is the following: one introduces two constants, $g_1$ and $g_2$, into $L$, 
so that the Lagrangian becomes
$L=\frac{1}{2}(\partial_\mu \phi \partial^\mu \phi - m^2 \phi^2) + g_1V(g_2\phi)$.
$g_2$ will be used to handle the problem that 
$V(\phi)$ and $V'(\phi)$ ($V'$ denoting the derivative of $V$) 
do not generally have simple scaling transformation properties under
$\delta_{\mathrm{sc}} \phi = \phi$. 
The Euler--Lagrange derivative of $L$ with respect to $\phi$ is
\beq
F=\frac{\delta L}{\delta \phi}= -\partial_\mu \partial^\mu\phi - m^2\phi + g_1g_2V'(g_2\phi)\, .
\eeq
$K^\mu = -h^\mu L$ can again be chosen in (\ref{eq.K}) for spacetime translations, and the Noether current $J^\mu_N$ can 
also be written in the form (\ref{eq.jn}). 

For applying the constructions of Section \ref{sec.ext}, one promotes both $g_1$ and $g_2$ to fields, and one defines
the scaling transformation to be
\beq
\delta_{\mathrm{sc}}\phi=\phi\, ,\qquad \delta_{\mathrm{sc}} g_1=2g_1\, ,\qquad \delta_{\mathrm{sc}} g_2=-g_2\, .
\eeq
This is a symmetry of the system $F=0$, $\partial_\mu g_1=0$, $\partial_\mu g_2=0$, since 
$\delta_{\mathrm{sc}} F = F$, $\delta_{\mathrm{sc}} (\partial_\mu g_1) = 2\partial_\mu g_1$, 
$\delta_{\mathrm{sc}} (\partial_\mu g_2) = -\partial_\mu g_2$. $J^\mu_N$ also transforms in the desired way, i.e.\ 
$\delta_{\mathrm{sc}} J^\mu_N = 2J^\mu_N$. 
We note that $\delta_{\mathrm{sc}} (g_2\phi)=0$, thus also 
$\delta_{\mathrm{sc}} V(g_2\phi)= \delta_{\mathrm{sc}} V'(g_2\phi) = 0$. The reason for introducing $g_2$ was 
precisely to achieve this, and thus to remedy the problem that
$V(\phi)$ and $V'(\phi)$ do not generally have simple scaling transformation properties under
$\delta_{\mathrm{sc}} \phi = \phi$. 

In the same way as in Section \ref{sec.ex2}, $q=\delta\phi=-h^\mu\partial_\mu\phi$ is a parameterized multiplier for $F$
and for the conserved current $J^\mu=-J^\mu_N$. 
$j^\mu_{(F\rho)}$ and $j^\mu_{(F\rho)}|_{\rho=q}$ are also given by the same expressions as in Section \ref{sec.ex2}, 
$\frac{\partial J^\mu}{\partial g_1} = -h^\mu V(g_2\phi)$ and 
$\frac{\partial J^\mu}{\partial g_2} = -h^\mu g_1\phi V'(g_2\phi)$,
thus (\ref{eq.jth}) gives   
\beq 
\label{eq.jres2}
j^\mu = (\partial^\mu\phi\, \partial_\lambda \phi - \phi\,\partial^\mu\partial_\lambda \phi)h^\lambda  
 -2g_1h^\mu V(g_2\phi) + g_1 g_2 h^\mu \phi V'(g_2\phi)\, .
\eeq
For $V(\phi)=\phi^n$ and $g_2=1$ this becomes identical with (\ref{eq.jres}).
$j^\mu_{(\tilde{F} q)}|_{\tilde{F}=F}$ is again $-Fh^\mu\phi$,  
and for $2J^\mu- j^\mu - j^\mu_{(\tilde{F} q)}|_{\tilde{F}=F}$ one obtains again (\ref{eq.jjj}).

As in Section \ref{sec.ex2}, the Lorentz symmetry of $L$ is not essential; the $x^0$-translation symmetric 
but not necessarily Lorentz symmetric generalization 
$L = \frac{1}{2}(\partial_\mu \phi \partial^\mu \phi - m^2 \phi^2) + g_1WV(g_2\phi)$, for instance, can be treated 
in the same way as the $W=1$ case above.
$J_N^\mu = - \partial^\mu\phi \partial_0\phi - \delta_0^\mu L$,
$\delta_{\mathrm{sc}}J^\mu_N = 2J^\mu_N$, and
the final result for $j^\mu$ is
$j^\mu = (\partial^\mu\phi\, \partial_0 \phi - \phi\,\partial^\mu\partial_0 \phi)  
-2g_1\delta_0^\mu W V(g_2\phi) + g_1 g_2 \delta_0^\mu \phi W V'(g_2\phi)$.
Furthermore, the same treatment can be extended without difficulty also 
to several fields of various kinds with a Lagrangian of the form $L=L_0[u]+L_I[u]$,
which would become 
\beq 
L=L_0[u]+g_1L_I[g_2u]
\eeq
after introducing $g_1$ and $g_2$,
where $L_0$ is quadratic and $L_I$ is a general local function.
For the symmetry it is again sufficient that $\delta u_i$ and $K^\mu$ be homogeneous under 
$\delta_{\mathrm{sc}}$ of order $\omega\ne -1$ and $1+\omega$, respectively. Under these conditions $J^\mu_N$ is also 
homogeneous of order $1+\omega$.

\subsubsection{Generalized Korteweg--de Vries equation}
\label{sec.ex4}

The generalized Korteweg--de Vries (gKdV) equation
is the partial differential equation
\beq
\label{eq.gkdv}
\partial_t u + u^p\partial_x u+\partial_x^3 u=0\quad\qquad (p>0)\, ,
\eeq
where $p$ is a parameter, $u(t,x)$ is the unknown function, and $t,x\in\RR$. 
In \cite{Anco2,AK} the following five multipliers were found for this equation:
\beq
q_1=1\, ,\qquad q_2=u\, ,\qquad q_3=\partial_x^2 u+\frac{1}{p+1}u^{p+1}\qquad (p>0)\, ,
\eeq
\beq
q_4=x-tu\qquad (p=1)\, ,
\eeq
\beq
q_5=t(3\partial_x^2 u+u^3)-xu\qquad (p=2)\, .
\eeq
The corresponding conserved currents (the components of which we denote by $J^t$ and $J^x$) were found to be
\beq
J_1^t \ = \ u\, , \qquad J_1^x \ = \ \frac{1}{p+1}u^{p+1}+\partial_x^2 u
\eeq
\beq
J_2^t \ = \ \frac{1}{2}u^2\, ,\qquad J_2^x \ = \ \frac{1}{p+2}u^{p+2}+u\partial_x^2 u-\frac{1}{2}(\partial_x u)^2
\eeq
\bea
J_3^t & = &\frac{1}{2}u\partial_x^2 u+\frac{1}{(p+1)(p+2)}u^{p+2}\, ,\\
J_3^x & = & \frac{1}{2(p+1)^2}u^{2p+2}+\frac{1}{p+1}u^{p+1}\partial_x^2 u + \frac{1}{2}((\partial_x^2 u)^2 +\partial_t u\partial_x u)-u\partial_{tx}u
\eea
\beq
J_4^t \ = \ xu-\frac{1}{2}tu^2\, ,\qquad
J_4^x \ = \ t\biggl(\frac{1}{2}(\partial_x u)^2-u\partial_x^2 u-\frac{1}{3}u^3\biggr)+x\biggl(\partial_x^2 u+\frac{1}{2}u^2\biggr)-\partial_x u
\eeq
\bea
J_5^t & = & \frac{1}{2}(3tu\partial_x^2 u -x u^2)+\frac{1}{4}tu^4\\
J_5^x & = & t\biggl(\frac{3}{2}((\partial_x^2 u)^2+\partial_t u\partial_x u) +u^3\partial_x^2 u-
\frac{3}{2}u\partial_{tx}u+\frac{1}{6}u^6\biggr)\nonumber\\
&& +x\biggl(\frac{1}{2}(\partial_x u)^2-u\partial_x^2 u-\frac{1}{4}u^4\biggr) - \frac{1}{2}u\partial_x u\, . 
\eea

For applying the constructions of Section \ref{sec.ext}, an additional constant parameter $g$ ($g>0$)
is introduced into the nonlinear part of (\ref{eq.gkdv}):
\beq
\label{eq.modgkdv}
F=\partial_t u + gu^p\partial_x u+\partial_x^3 u\, .
\eeq 
$g$ is also introduced into the multipliers and into the corresponding conserved currents:
$q_1$ and $q_2$ are not changed, 
$q_3=\partial_x^2 u+\frac{1}{p+1}gu^{p+1}$, $q_4=x-tgu$, 
$q_5=t(3\partial_x^2 u+gu^3)-xu$,
and the currents are modified in the following way: 
$u^{p+1}\to g u^{p+1}$ in $J_1^x$;
$u^{p+2}\to g u^{p+2}$ in $J_2^x$ and $J_3^t$;
$u^{2p+2}\to g^2 u^{2p+2}$ and $u^{p+1}\to g u^{p+1}$ in $J_3^x$;
in $J_4^t$ and $J_4^x$ the terms quadratic in $u$ (including the derivatives of $u$) are multiplied by $g$
and the third order term is multiplied by $g^2$;
$u^{4}\to g u^{4}$ in $J_5^t$;
in $J_5^x$ the terms that are fourth order in $u$ (including the derivatives of $u$) are multiplied by $g$
and the sixth order term is multiplied by $g^2$. 
By these modifications $q_1,\dots, q_5$ become parameterized multipliers. 
We note that a constant parameter was introduced into 
the gKdV equation and into the multipliers in the same way in \cite{Anco2} in the context of the problem of finding the 
conserved currents that correspond to the multipliers. In \cite{Anco2} the parameter was denoted by $\mu$.
It is also remarkable that in the examples in the previous two subsections the multipliers do not depend on the $g$ constants,
thus the dependence of $q_3$, $q_4$ and $q_5$ on $g$ is a new feature of the present example. 

For the extended DE system one takes $F=0$, $\partial_\mu g=0$, and 
the scaling transformation is chosen to be
\beq
\label{eq.scgkdv}
\delta_{\mathrm{sc}}u=u\, ,\qquad \delta_{\mathrm{sc}} g=-pg\, .
\eeq
This scaling transformation was also considered in \cite{Anco2}.
$\delta_{\mathrm{sc}} (gu^p)=0$, thus $\delta_{\mathrm{sc}} F = F$, 
and (\ref{eq.scgkdv}) is a symmetry of the extended system of DEs.
Furthermore, the multipliers and the conserved currents are homogeneous under (\ref{eq.scgkdv}):
\bea 
\label{eq.ni0}
&& \delta_{\mathrm{sc}} q_i = \kappa_i q_i \quad (i=1,\dots,5)\, ; \qquad \kappa_1=\kappa_4=0,\quad \kappa_2=\kappa_3=\kappa_5=1\, ,\\
\label{eq.ni}
&&\delta_{\mathrm{sc}} J^\mu_i = \omega_i J^\mu_i\, ; \hspace{2.75cm} \omega_1=\omega_4=1,\quad \omega_2=\omega_3=\omega_5=2\, .
\eea
By theorem \ref{th.2}, the currents given by (\ref{eq.jth}) are thus equivalent with
$J^\mu_1$, $2J^\mu_2$, $2J^\mu_3$, $J^\mu_4$, $2J^\mu_5$, respectively.
For $j^\mu_{(F\rho)}$ one finds
\beq
j_{(F\rho)}^t=\rho u,\qquad 
j_{(F\rho)}^x=g\rho u^{p+1}+\rho\partial_x^2 u-\partial_x\rho\,\partial_x u+\partial_x^2\rho\, u\, .
\eeq
In the first two cases $j^\mu$, given by (\ref{eq.jth}), is exactly $J^\mu_1$ and $2J^\mu_2$; $j^\mu_{(\tilde{F}q)}=0$, 
$\frac{\partial J^\mu_1}{\partial g} = (0,\frac{1}{1+p}u^{p+1})$,
$\frac{\partial J^\mu_2}{\partial g} = (0,\frac{1}{2+p}u^{p+2})$.
In the third case, $\frac{\partial J^\mu_3}{\partial g} = (\frac{1}{(1+p)(2+p)}u^{p+2},
\frac{g}{(1+p)^2}u^{2p+2}+\frac{1}{1+p}u^{p+1}\partial_x^2 u)$,
$j_{(\tilde{F} q)}|_{\tilde{F}=F}=(0,F\partial_x u - u D_x F)$,
and $2J^\mu_3 = j^\mu + j^\mu_{(\tilde{F} q)}|_{\tilde{F}=F}$. 
In the fourth case, $\frac{\partial J^\mu_4}{\partial g}=
(-\frac{1}{2}tu^2,t(\frac{1}{2}(\partial_x u)^2-u\partial_x^2 u )
+\frac{1}{2}x u^2-\frac{2}{3}gu^3)$, $j^\mu_{(\tilde{F} q)}=0$, and $j^\mu=J^\mu_4$.
In the fifth case,  $\frac{\partial J^\mu_5}{\partial g}=
(\frac{1}{4}tu^4, tu^3\partial_x^2 u + \frac{1}{3}gu^6-\frac{1}{4}xu^4)$,
$j^\mu_{(\tilde{F} q)}|_{\tilde{F}=F}=(0,3tF\partial_x u-3tu D_x F)$,
and $2J^\mu_5 = j^\mu + j^\mu_{(\tilde{F} q)}|_{\tilde{F}=F}$. 
On the solutions of the extended system $j^\mu$ is equal to $\omega_iJ^\mu_i$ (see (\ref{eq.ni}) for $\omega_i$) in all cases.

The results of \cite{Anco2,AK} on the symmetries and conservation laws of the gKdV equation  
show that the gKdV equation has a scaling symmetry even in its original form, 
and the conserved currents $J^\mu_1, \dots, J^\mu_5$ are homogeneous under its action
up to equivalence. This means that $J^\mu_1, \dots, J^\mu_5$ can also be reproduced (up to equivalence) 
by applying the embedding method to the original gKdV equation, 
if the scaling weights of $J^\mu_1, \dots, J^\mu_5$ with respect to the scaling symmetry mentioned in
\cite{Anco2,AK} are not zero.
However, the scaling weights of $J^\mu_4$ and $J^\mu_5$ are zero, and there are special values of $p$
for which the scaling weight of $J^\mu_1$ or $J^\mu_2$ also becomes zero. 
On the other hand, the extended system discussed above with the scaling symmetry (\ref{eq.scgkdv}) is free from this difficulty.

The example of the gKdV equation also admits a generalization.
Let us assume that $F^a[u]$, a multiplier $q_a[u]$, and the corresponding conserved current $J^\mu[u]$ take the form 
\beq 
\label{eq.h1}
F^a[u]=\sum_k F^a_k[u]\, ,\qquad q_a[u]=\sum_m q_{ma}[u]\, ,\qquad J^\mu[u] = \sum_n J_n^\mu[u]\, ,
\eeq
where $F^a_k[u]$, $q_{ma}[u]$, $J_n^\mu[u]$ are homogeneous of order $\alpha_k$, $\beta_m$, $\gamma_n$,
respectively, under the scaling transformation $\delta_{\mathrm{sc}}u_i = u_i$. The dimension of the base manifold 
can be arbitrary and $u_i$ is also not restricted to be a single scalar field.
In this case one can introduce the constant parameter $g$ ($g>0$) into $F^a$, $q_a$, $J^\mu$ by taking
\beq
\label{eq.h2}
F^a[u,g]=\sum_k g^{\eta-\alpha_k} F^a_k[u]\, ,\qquad q_a[u,g]=\sum_m g^{\rho-\beta_m}q_{ma}[u]\, ,
\qquad J^\mu[u,g] = \sum_n g^{\eta+\rho-\gamma_n} J_n^\mu[u]\, ,
\eeq
where $\eta$ and $\rho$ are arbitrary real numbers for which $\eta+\rho\ne 0$. 
In this way $q_a$ becomes a parameterized multiplier for $F^a$ and $J^\mu$. Then one promotes $g$ to a field with scaling transformation property
$\delta_{\mathrm{sc}}g = g$.    
$\delta_{\mathrm{sc}} (g^{\eta-\alpha_k} F^a_k[u]) = \eta g^{\eta-\alpha_k} F^a_k[u]$, thus
$\delta_{\mathrm{sc}} F^a = \eta F^a$, 
therefore $\delta_{\mathrm{sc}}$ 
is a symmetry of the extended DE system $F^a=0$, $\partial_\mu g = 0$. 
Furthermore, $q_a$ and $J^\mu$ are also homogeneous under
$\delta_{\mathrm{sc}}$: $\delta_{\mathrm{sc}} q_a = \rho q_a$, $\delta_{\mathrm{sc}} J^\mu= (\eta+\rho)J^\mu$,
thus the current (\ref{eq.jth}) is equivalent with $(\eta+\rho)J^\mu$.

In the case of the gKdV equation 
one can choose $F_1=\partial_t u+\partial_x^3 u$, $F_2=u^p\partial_x u$, 
for which $\alpha_1=1$, $\alpha_2=1+p$, and $q_i$ and $J_i^\mu$, $i=1,\dots,5$, obviously also have the form prescribed in (\ref{eq.h1}).
For $\eta_i$ one can take $\eta_i=1$, and then $F$ becomes 
$\partial_t u + g^{-p}u^p\partial_x u + \partial_x^3 u$ after introducing $g$. 
For $\rho_i$ one can take $\rho_1=\rho_4=0$, $\rho_2=\rho_3=\rho_5=1$, and then
$q_i$ and $J^\mu_i$ also take the form given previously in this subsection, but with $g\to g^{-p}$, like $F$.
The previous formulas for $F$, $q_i$, $J^\mu_i$ can thus obviously be recovered by the replacement $g^{-p}\to g$,
which also leads to the transformation property (\ref{eq.scgkdv}) for $g$.

\section{Conclusion}
\label{sec.concl}

According to recent results, the multiplier method and the embedding method
differ significantly in the range of conserved currents they can generate.
Specifically, the embedding method was found to be generally less powerful in this respect than the multiplier method \cite{Anco1}.

With the aim of continuing the investigation of the relations
between the multiplier and the embedding methods and improving on the result mentioned above,
we studied simple extended forms (\ref{eq.deext}) of general DE systems, obtained by promoting constant parameters
of the DEs to dependent variables.
We derived a variant of a known fundamental result about the connection between the two methods
for the extended DEs,  
and showed that, up to equivalence, the multipliers of an extended DE system
consist of the parametric multipliers of the 
original system accompanied by the derivatives of the corresponding conserved currents with respect to the parameters.
In addition, we noted that the extended DE systems have adjoint-symmetries composed of the conserved currents of the extended system,
which can be used in the embedding method together with symmetries that act nontrivially 
on the parameters (see Section \ref{sec.as}).

In Section \ref{sec.f} we studied the simplest extension of arbitrary systems of DEs (\ref{eq.gf}) and found that
by applying the embedding method to this extended system
it is possible to generate all local conserved currents of the original system,
moreover those conserved currents 
that correspond to a multiplier can be generated using an adjoint-symmetry that is also a multiplier.
In Section \ref{sec.examples} we discussed examples of other possible extensions 
that are more specific to the features of the DEs under consideration.
Although the titles of the subsections \ref{sec.ex2}, \ref{sec.ex3}, \ref{sec.ex4} indicate relatively special differential equations, 
generalizations of these equations to wide classes of DE systems, which include many equations of interest in physics,
were also given in these subsections.
The examples in Section \ref{sec.examples}, the results for (\ref{eq.gf}), and the existence of the adjoint-symmetries 
mentioned in Section \ref{sec.as}
show that the embedding method becomes significantly stronger
if it is also allowed to be applied to the extended forms of the original system of DEs.

In principle, the results for (\ref{eq.gf}) mean that taking into account the extended DE systems 
strengthens the embedding method to the maximal possible extent and solves 
the problem of the relative weakness of the embedding method. 
However, (\ref{eq.gf}) is a trivial extension
that does not give much help for finding conserved currents in practice, therefore it is important to consider
other extensions that make more use of the features of the DEs that one studies.
Although such extensions were discussed in Section \ref{sec.examples},  
it would still be interesting to explore the various possibilities more completely
and to investigate their practical usefulness in comparison with the multiplier method. 

A further point that is worth noting for the assessment of the virtues of the two methods is that  
the embedding method is also suitable for generating conserved currents associated with the symmetries of 
(\ref{eq.de}) without using any adjoint-symmetry. 
These currents are generally local conserved currents of the Euler--Lagrange equation system in which 
(\ref{eq.de}) is embedded, rather than of (\ref{eq.de}), nevertheless 
they are useful for some purposes---for example, for verifying approximate solutions of (\ref{eq.de})  
obtained by a numerical method. 

Finally we recall, regarding the relations between the two methods, that from the point of view of
the auxiliary Lagrangian (\ref{eq.l}) used in the embedding method the multiplier method
is a special case of Noether's theorem (see Theorem \ref{th.emb2}),
thus (\ref{eq.l}) underlies not only the embedding method but the multiplier method as well.

\section*{Acknowledgments}

The author is supported by the NKFIH grant no.\ K116505.

\appendix

\renewcommand{\theequation}{\Alph{section}.\arabic{equation}} 
\setcounter{equation}{0}

\section{Symmetries of Lagrangians}
\label{sec.noether}

In this appendix the continuous symmetries of Lagrangians and the corresponding conserved currents are discussed, 
after certain auxiliary formulas. Some of the notation used below is introduced in Section \ref{sec.prel}. 

\begin{prop}
\label{pr.a1}
Let $G$ be a homogeneous linear local function of $\epsilon^\alpha(x)$, i.e.
\beq
\label{eq.g1}
G= G_\alpha \epsilon^\alpha + G_\alpha^{\mu_1} \partial_{\mu_1} \epsilon^\alpha  
+  G_\alpha^{\mu_1\mu_2} \partial_{\mu_1\mu_2} \epsilon^\alpha + \dots\, ,
\eeq
where $G_\alpha^{\mu_1\mu_2}(x)$,  $G_\alpha^{\mu_1\mu_2\mu_3}(x)$, ... 
are completely symmetric in the upper indices, and the sum on the right hand side contains only finitely many terms.
$\epsilon^\alpha$ and $G_\alpha$, $G_\alpha^{\mu_1}$, $G_\alpha^{\mu_1\mu_2}$, ... may be anticommuting for some values of $\alpha$.
Under these conditions $G$ can be written as
\beq
\label{eq.g2}
G =  \hat{G}_\alpha \epsilon^\alpha + \partial_\mu \mc{G}^\mu\, ,
\eeq
where 
\beq
\label{eq.g3}
\hat{G}_\alpha =G_\alpha -\partial_{\mu_1} G^{\mu_1}_\alpha +\partial_{\mu_1\mu_2} G^{\mu_1\mu_2}_\alpha - \dots
\eeq
and
\beq
\label{eq.g4}
\mc{G}^\mu= G_\alpha^{\mu}  \epsilon^\alpha  +
(G_\alpha^{\mu\lambda_1} \partial_{\lambda_1} \epsilon^\alpha  - \partial_{\lambda_1}  G_\alpha^{\mu\lambda_1} \epsilon^\alpha)
+( G_\alpha^{\mu\lambda_1\lambda_2} \partial_{\lambda_1\lambda_2} \epsilon^\alpha - 
\partial_{\lambda_2} G_\alpha^{\mu\lambda_1\lambda_2} \partial_{\lambda_1} \epsilon^\alpha 
+ \partial_{\lambda_1\lambda_2} G_\alpha^{\mu\lambda_1\lambda_2} \epsilon^\alpha)+\dots\, .
\eeq
\end{prop}
(\ref{eq.g2}) can be derived by applying the differentiation rule of products in a straightforward way.

The first order variation of a Lagrangian density function $L[u]$ under 
a one-parameter transformation $u_i\to T_i\{\varsigma,u\}$ is defined as
$\delta L[u] = \frac{dL[u+\varsigma\delta u [u]]}{d\varsigma}|_{\varsigma=0}$;
clearly
\beq
\delta L  =  \frac{\partial L}{\partial u_i}\delta u_i +
\frac{\partial L}{\partial (\partial_\mu u_i)}D_\mu\delta u_i +
\frac{\partial L}{\partial (\partial_{\mu\nu}u_i)}D_{\mu\nu} \delta u_i + \dots\, . 
\label{eq.dl1} 
\eeq
By applying the rule of differentiation of products (in particular (\ref{eq.g1})-(\ref{eq.g4}))
to the right hand side of (\ref{eq.dl1}), 
$\delta L$ can be rewritten as
\beq
\label{eq.dl3}
\delta L[u,\delta u[u]] =  \mb{E}^i(L)[u]\delta u_i[u] 
+ D_\mu j^\mu[u,\delta u[u]]\, , 
\eeq
where $\mb{E}^i(L)[u]$ is the Euler--Lagrange derivative given by (\ref{eq.el1}) and
\bea
&& \hspace{-1.5cm} j^\mu  =   \frac{\partial L}{\partial (\partial_\mu u_i)}\delta u_i+
\biggl(\frac{\partial L}{\partial (\partial_{\mu\nu}u_i)}D_\nu \delta u_i
-D_\nu\frac{\partial L}{\partial (\partial_{\mu\nu}u_i)} \delta u_i\biggr)\nonumber\\
&& + \biggl(\frac{\partial L}{\partial (\partial_{\mu\nu\lambda}u_i)}D_{\nu\lambda}\delta u_i
-D_\nu\frac{\partial L}{\partial (\partial_{\mu\nu\lambda}u_i)}D_\lambda\delta u_i
+D_{\nu\lambda}\frac{\partial L}{\partial (\partial_{\mu\nu\lambda}u_i)}\delta u_i\biggr)+\dots\, .
\label{eq.elc}
\eea

If
\beq
\label{eq.K}
\delta L= D_\mu K^\mu 
\eeq
holds with some $K^\mu[u]$ for arbitrary configurations of $u_i$,   
then $u_i\to u_i+\varsigma\,\delta u_i$ is called a \emph{symmetry of $L$}
and from (\ref{eq.dl3}) it follows that 
\beq
\label{eq.n1}
D_\mu J^\mu + \mb{E}^i \delta u_i
= 0\, ,
\eeq
where $J^\mu$ is defined as
\beq
\label{eq.n2}
J^\mu = j^\mu - K^\mu\, . 
\eeq
$J^\mu$ is called the \emph{Noether current} associated with the symmetry $u_i\to u_i+\varsigma\,\delta u_i$.
Since $K^\mu$ is determined by (\ref{eq.K}) up to adding identically conserved currents, 
$J^\mu$ is also determined only up to identically conserved currents.   
If $u_i$ also satisfy their Euler--Lagrange equations, 
i.e.\ $\mb{E}^i(L)[u]=0$,
then from (\ref{eq.n1}) it follows that
$J^\mu$ is conserved. This result, together with its converse, which we do not discuss here, is known as Noether's theorem.
For more detailed expositions of Noether's theorem the reader is referred to \cite{Olver,BCA,Noether,Y}.

\subsection{On-shell symmetries}
\label{sec.oss}

For the conservation of $J^\mu$ it is sufficient that the symmetry condition (\ref{eq.K}) holds only 
on the solutions of the Euler--Lagrange equations. In this case
we call $u_i\to u_i+\varsigma\,\delta u_i$ an \emph{on-shell symmetry of $L$}.
$K^\mu$ and thus also $J^\mu$ are rather undetermined in the case of on-shell symmetries,
since any conserved current can be added to $K^\mu$. Moreover, (\ref{eq.K}) is obviously satisfied 
on the solutions of the Euler--Lagrange equations for any transformation if $K^\mu=j^\mu$ is chosen, 
although with this choice $J^\mu=0$.
It is therefore essential to restrict $K^\mu$ in a suitable way if one considers on-shell symmetries.
For the present paper, for instance, $K^\mu=0$ is suitable.


\end{document}